\documentclass[preprint,review,12pt]{elsarticle}

\usepackage{amssymb}
\usepackage{amsmath,bm}
\usepackage{colortbl,booktabs}
\usepackage{tabularx}
\usepackage{threeparttable}
\usepackage{multicol,multirow}
\usepackage{subfigure}
\usepackage [font=footnotesize,labelfont=bf]{caption}
\usepackage{rotating}
\usepackage{tikz}
\usepackage{hyperref}
\usepackage{color}
\definecolor{R}{rgb}{1, 0, 0}
\definecolor{G}{rgb}{0, 0.5, 0}
\definecolor{B}{rgb}{0, 0, 1}

\definecolor{yjzhuCol}{rgb}{0.9, 0, 0.2}
\definecolor{czhangCol}{rgb}{0.9, 0, 0.2}
\definecolor{xyhuCol}{rgb}{0.9, 0, 0.9}

\journal{Elsevier}

\begin{document}
\captionsetup[figure]{labelfont={bf},labelformat={default},labelsep=period,name={Fig.}}
\begin{frontmatter}

\title{Essentially non-hourglass and non-tensile-instability SPH elastic dynamics}

%\tnotetext[label0]{This is only an example}

\author[address1]{Shuaihao Zhang}
\ead{szhang07@connect.hku.hk}
\author[address1]{Sérgio D.N. Lourenço}
\ead{lourenco@hku.hk}
\author[address2]{Dong Wu}
\ead{dong.wu@tum.de}
\author[address2]{Chi Zhang}
\ead{c.zhang@tum.de}
\author[address2]{Xiangyu Hu \corref{mycorrespondingauthor}}
\cortext[mycorrespondingauthor]{Corresponding author.}
\ead{xiangyu.hu@tum.de}
\address[address1]{Department of Civil Engineering, The University of Hong Kong, Pokfulam, Hong Kong SAR, China}
\address[address2]{School of Engineering and Design, Technical University of Munich, 85748 Garching, Germany}

\begin{abstract}
	
Since the tension instability was discovered in updated Lagrangian smoothed particle hydrodynamics (ULSPH) at the end of the 20th century, 
researchers have made considerable efforts to suppress its occurrence. However, up to the present day, this problem has not been fundamentally resolved.
In this paper, the concept of hourglass modes is firstly introduced into ULSPH, and the inherent causes of tension instability in elastic dynamics are clarified based on this brand-new perspective. 
Specifically, we present an essentially non-hourglass formulation by decomposing the shear acceleration with the Laplacian operator, and a comprehensive set of challenging benchmark cases for elastic dynamics is used to showcase that our method can completely eliminate tensile instability by resolving hourglass modes. 
The present results reveal the true origin of tension instability and challenge the traditional understanding of its sources, i.e., hourglass modes are the real culprit behind inducing this instability in tension zones rather that the tension itself.
Furthermore, a time integration scheme known as dual-criteria time stepping is adopted into the simulation of solids for the first time, to significantly enhance computational efficiency.

\end{abstract}


\end{frontmatter}
% \linenumbers
%%%%%%%%%%%%%%%%%%%%%%%%%%%%%%%%%%%%%%%%%%%%%%%%%%%%%%%%%%%%%
%
% 1 Introduction
%
%%%%%%%%%%%%%%%%%%%%%%%%%%%%%%%%%%%%%%%%%%%%%%%%%%%%%%%%%%%%%
\section{Introduction}
\label{introduction}
Smoothed particle hydrodynamics (SPH), 
original proposed by Lucy \cite{lucy1977numerical} and Gingold 
and Monaghan \cite{gingold1977smoothed} for simulating astrophysical problems, 
is a fully Lagrangian particle-based method. 
In SPH, the physical quantities such velocity, position, and stress are carried by each particle, 
and the motion of particles is described in the Lagrangian framework, 
which is naturally well-suited for simulating problems involving large deformations, 
especially fracture and failure. 
Over the past 40 years, 
SPH has been extensively developed and improved, 
and it has been successfully applied to simulate various physical problems 
including fluid dynamics \cite{morris1997modeling, hu2006multi}, 
solid dynamics \cite{gray2001sph, johnson1996sph}, 
and fluid-solid interactions \cite{antoci2007numerical, khayyer20213d}. 

Based on whether the particle configurations,
which define the neighbors of each particle, 
are updated during the simulation, 
the SPH methods for solid dynamics 
can be classified into the total Lagrangian SPH (TLSPH) \cite{vignjevic2006sph} 
and updated Lagrangian SPH (ULSPH) \cite{gray2001sph, monaghan2000sph}. 
TLSPH is able to handle the elastic and plastic dynamics efficiently 
as it saves the time required for updating particle configurations.
Compared with TLSPH, while ULSPH is able to cope with 
material failure and fracture beyond elastic or plastic deformations more naturally
with updating particle configurations at each time step, 
it faces two important drawbacks associated 
with elastic dynamics:
one is the persistent issue of tensile instability;
the other is poorer efficiency 
due to the computational effort and memory latency 
when the particle configurations are updated frequently.  

\begin{figure}[tb!]
	\centering
	\includegraphics[trim = 0cm 0cm 0cm 0cm, clip,width=.85\textwidth]{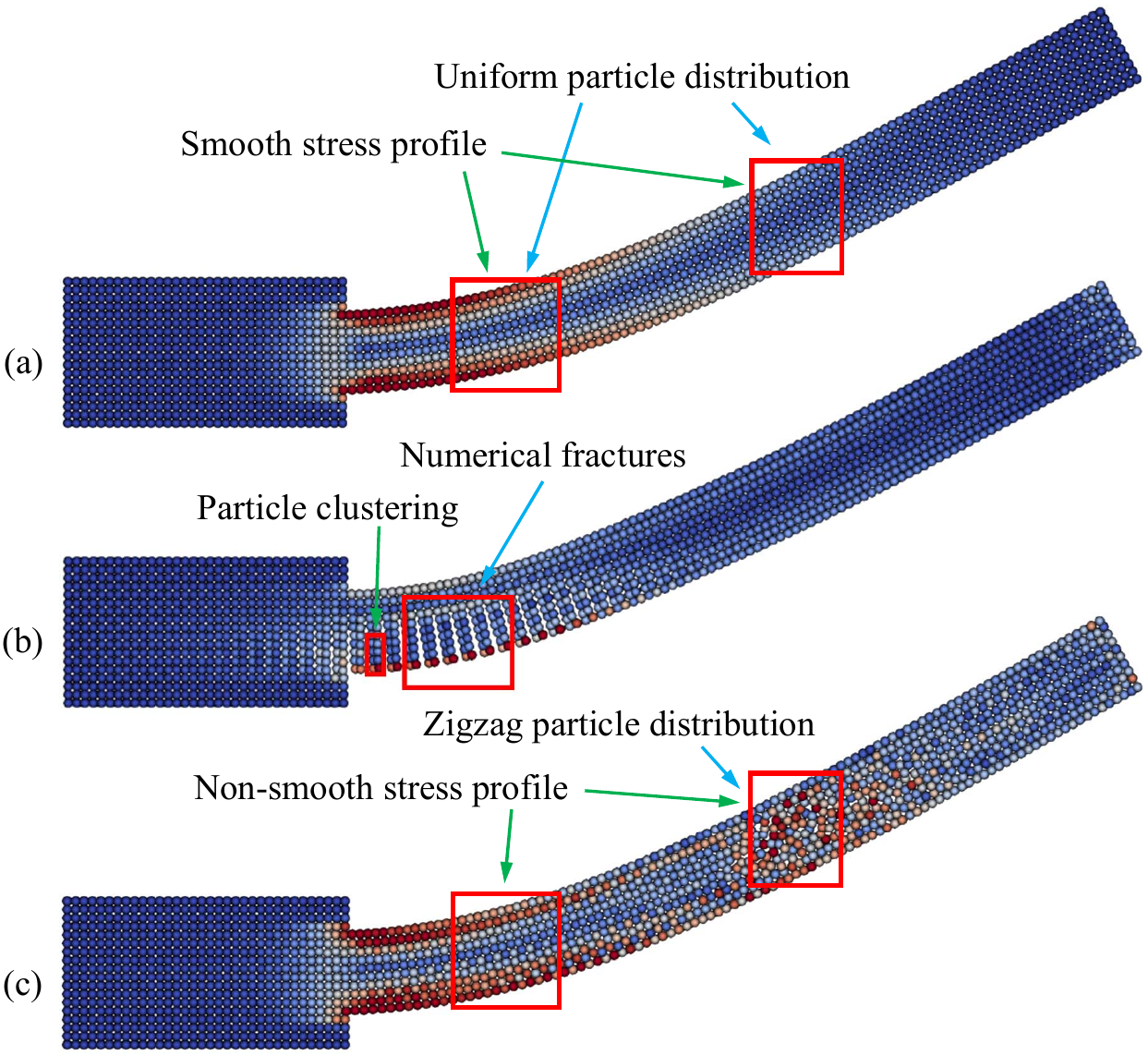}
	\caption{Illustration for (a) non-tensile-instability and non-hourglass modes, (b) tensile instability and (c) hourglass modes in ULSPH simulations of 2D oscillating plates. The particles are colored with von Mises stress.}
	\label{figs:hourglass-tensile-instability-illustration}
\end{figure}

As shown in Fig. \ref{figs:hourglass-tensile-instability-illustration}b,
tensile instability in ULSPH is often associated 
with particle clustering and numerical/artificial fractures,
and was first studied
by Swegle et al. \cite{swegle1995smoothed} in 1995. 
At that time, it was believed to be caused by tension stress.
Since then, different approaches have been proposed to address this problem 
as reported in the literatures 
\cite{randles1996smoothed, johnson1996normalized, dyka1997stress}. 
However, these methods also have their own issues, 
such as failure to maintain conservation properties, 
low computational efficiency, or limited applicability, 
being only suitable for specific cases rather than universally applicable \cite{mandell1996computational, dilts1999moving, randles2000normalized}.
Later, in 2000, 
inspired by the repulsive interactions observed among closely spaced atoms, 
Monaghan \cite{monaghan2000sph} introduced a small repulsive force 
(named the artificial stress) between particles in SPH 
to prevent particle clustering and then remove tensile instability. 
Based on the artificial stress and the signs of principal stresses, 
Gray et al. \cite{gray2001sph} further developed this approach 
by determining the parameters in artificial stress from 
the dispersion relation for elastic waves.

Despite the broader recognition 
compared to many other approaches in addressing tension instability, 
the artificial stress method still faces the following limitations: 
(1) it has two parameters requiring case-dependent tuning; 
(2) it may fail in scenarios where the deformation is significant or 
when dealing with materials featuring high Poisson's ratio 
\cite{zhang2017generalized, lobovsky2007smoothed};
(3) its extension for three-dimensional (3D) simulations is yet to be developed. 
The third issue is due to the fact 
that the derivation of the artificial stress term 
is based on a two-dimensional (2D) scenario \cite{gray2001sph}. 
To our best knowledge, 
there are no documented instances utilizing 
the artificial stress in 3D simulations. 

Tensile instability is also found in SPH simulations of fluids, 
in which the negative pressure, especially in vortical flows,
leads to the generation of artificial void regions. 
A commonly used strategy is applying a constant background pressure \cite{litvinov2015towards} to keep positive pressure everywhere in 
the simulations of incompressible flow 
with the weakly compressible SPH (WCSPH) method.
Although tensile instability is first found for SPH elastic dynamics, 
its generally effective remedies are first found for SPH fluid dynamics     
\cite{hu2007incompressible, xu2009accuracy, adami2013transport}.
Further analysis \cite{litvinov2015towards} 
suggested that tensile stability in SPH fluid dynamics is 
highly relevant to zero-order consistency error, 
very often due to the non-regular particle distributions 
which are typically generated by the complex velocity gradient in flow field. 

Such observations in flow simulations lead to 
a puzzle for ULSPH in solid dynamics,
where the velocity gradient is much more regular than that of a flow field
and theoretically should only generate very regular particle distributions
hence no tensile instability,
except when very large deformation or material failure happens.
On the other hand, 
it is well known that TLSPH elastic dynamics does not suffer from 
tensile instability as the ULSPH counterpart.
Such property, however, can be well explained by 
the non-updated configuration obtained from  
the very regular initial particle distribution.

In this work, we propose to address these above-mentioned two drawbacks of ULSPH in elastic dynamics. 
Firstly, we solve the puzzle 
why the supposed very regular velocity field of elastic dynamics still 
suffers tensile instability.
For this, other than tension,
we relate such issue with the hourglass modes and shear stress,
as shown in Fig. \ref{figs:hourglass-tensile-instability-illustration}c,
a numerical instability producing "zigzag" particle distribution initially
found in TLSPH elastic dynamics when the deformation is very large \cite{ganzenmuller2015hourglass}.
Specifically, we argue that, 
at least for the elastic dynamics without material failure,
the previous consensus of "tensile instability" in ULSPH is 
actually not caused by tension  
but hourglass modes introduced by the nested formulation on 
the acceleration from shear stress or 
the collocation of deformation and stress at the same particle positions, 
just like it has been recently found in TLSPH \cite{wu2023essentially}. 

Based on this argument, 
we develop a non-nested angular-momentum conservative ULSPH formulation 
for computing the shear stress induced acceleration 
and essentially eliminate the hourglass modes.
We show that the new formulation is tensile stable,
i.e. without tensile instability, 
even when very large tension and deformation are involved.
Different from Ref. \cite{gray2001sph}, 
the present solution works for both 2D and 3D scenarios 
without resorting to case-dependent tuning. 

Secondly, to improve computational efficiency, 
a dual-criteria time stepping method \cite{zhang2020dual} 
is incorporated into ULSPH simulations of elastic dynamics for the first time. 
There are two time steps named the advection time step
and the much smaller acoustic time step. 
By updating of particle configurations only in the advection time step, 
the frequency for updating particle configurations can be minimized, 
and the calculation time can be reduced. 
It is worth mentioning that the speed of sound in solid simulations 
is the true sound speed, much higher than 
the artificial sound speed used in the WCSPH method for fluids. 
This implies the dual-criteria time stepping scheme
leads to a significant enhancement in computational 
efficiency for solid simulations.

The remainder of this article is arranged as follows.
The basic theory of elastic dynamics is introduced in Section \ref{governing-equation}.
The original formulation and the present essentially non-hourglass and non-tensile-instability formulation for elastic dynamics are described in Section \ref{original-SPH-formulation} and Section \ref{non-hourglass-formulation-section} respectively.
The dual-criteria time stepping scheme for solid simulations is described in Section \ref{time-integration}, and a set of benchmark cases for elastic dynamics are then used to validate the convergence, accuracy and stability of the proposed method in Section \ref{validation}.
Section \ref{conclusions} draws the conclusion.
For future in-depth research, all the code used in this study has been open-sourced in the SPHinXsys repository \cite{zhang2021sphinxsys} at https://www.sphinxsys.org and https://github.com/Xiangyu-Hu/SPHinXsys.

%%%%%%%%%%%%%%%%%%%%%%%%%%%%%%%%%%%%%%%%%%%%%%%%%%%%%%%%%%%%%
%
% 2 Governing equations and constitutive relations
%
%%%%%%%%%%%%%%%%%%%%%%%%%%%%%%%%%%%%%%%%%%%%%%%%%%%%%%%%%%%%%
\section{Governing equations and constitutive relations}
\label{governing-equation}

In a Lagrangian framework, the governing equations include mass and momentum conservation for continuum mechanics are defined as
\begin{equation}
    \frac{\text{d} \rho }{\text{d} t} = -\mathbf \rho \nabla \cdot \mathbf v
   \label{continuity-equation}
\end{equation}
\begin{equation}
    \frac{\text{d} \mathbf v}{\text{d} t} = \frac{1}{\mathbf \rho}\nabla \cdot \bm{\sigma} + \mathbf g
   \label{momentum-equation}
\end{equation}
where ${\rho}$ is the density, $\mathbf{v}$ is velocity, ${t}$ is the time, $\bm{\sigma}$ is the stress tensor, and $\mathbf{g}$ is the body force. The total stress tensor $\bm{\sigma}$ can be divided into two terms, i.e., the hydrostatic pressure and the shear stress, as shown below.
\begin{equation}
    \bm{\sigma} = -p \mathbf I + \bm{\sigma}^s
   \label{pressure-shear-stress}
\end{equation}
where ${p}$ is the hydrostatic pressure, $\mathbf{I}$ is the identity matrix, and $\bm{\sigma}^s$ is the shear stress. The pressure ${p}$ can be evaluated from density based on an artificial equation of state \cite{gray2001sph}.
\begin{equation}
    p=c_0^2(\rho - \rho_0)
   \label{EOS}
\end{equation}
where ${\rho_0}$ and ${\rho}$ are the initial and the current density respectively. ${c_0}$ is the sound speed, which is expressed as \cite{zhang2017generalized}
\begin{equation}
    c_0=\sqrt{\frac{E}{3(1-2\nu)\rho_0}} 
   \label{sound-speed}
\end{equation}
where ${E}$ is the Young's modulus, ${\nu}$ is the Poisson's ratio of the given material.
The shear stress is the integral of the shear stress rate with respect to time. 
\begin{equation}
    \bm{\sigma}^s=\int_{0}^{t} \dot{\bm{\sigma}}^s  \text{d}t 
   \label{shear-stress-integral}
\end{equation}

For a linear elastic model, the shear stress rate is defined as
\begin{equation}
    \dot{\bm{\sigma}}^s = 2G( \dot{\bm \varepsilon}  -\frac{1}{d}tr(\dot{\bm \varepsilon})\mathbf I)
   \label{stress-rate}
\end{equation}
where G is the shear modulus. ${tr(\varphi  )}$ indicates the trace of a variable ${\varphi}$ and ${\dot{\varphi}}$ is the change rate with time for the variable ${\varphi}$ (${\varphi}$ is an arbitrary variable). ${d}$ represents the space dimension, and ${d=2}$ and ${3}$ for 2D and 3D cases respectively. ${\dot{\bm \varepsilon}}$ is strain rate, which is defined as
\begin{equation}
    \dot{\bm \varepsilon} =\frac{1}{2}(\nabla \mathbf v + (\nabla \mathbf v)^T)
   \label{strain-rate}
\end{equation}
where ${\nabla \mathbf v}$ donates the velocity gradient,
and superscript $T$ indicates the transpose of a tensor.

%%%%%%%%%%%%%%%%%%%%%%%%%%%%%%%%%%%%%%%%%%%%%%%%%%%%%%%%%%%%%
%
% 3 Original SPH formulation
%
%%%%%%%%%%%%%%%%%%%%%%%%%%%%%%%%%%%%%%%%%%%%%%%%%%%%%%%%%%%%%
\section{Original SPH formulation}
\label{original-SPH-formulation}
According to Eq. \eqref{momentum-equation} and Eq. \eqref{pressure-shear-stress}, The acceleration related to volumetric (hydrostatic pressure) and deviatoric part (shear stress) of the stress tensor can be express as
\begin{equation}
    \dot{\mathbf v}^p=-\frac{1}{\mathbf \rho}\nabla p
   \label{normal-acc}
\end{equation}
\begin{equation}
    \dot{\mathbf v}^s=\frac{1}{\mathbf \rho}\nabla \cdot \bm{\sigma}^s
   \label{shear-acc}
\end{equation}
where $\dot{\mathbf v}^p$ and $\dot{\mathbf v}^s$ donate the velocity change rate (acceleration) induced by hydrostatic pressure and shear stress respectively. Then the total velocity change rate $\dot{\mathbf v}$=$\dot{\mathbf v}^p$+$\dot{\mathbf v}^s$+$\mathbf{g}$. 

A low-dissipation Riemann solver \cite{zhang2017weakly} is incorporated in the WCSPH to discrete the continuity equation and the momentum equation for hydrostatic pressure. 
\begin{equation}
    \frac{\text{d} \rho_i }{\text{d} t} = 2 \rho_i \sum_{j} \frac{m_j}{\rho_j} (U^*-\mathbf v_{ij} \mathbf e_{ij})  \frac{\partial  W_{ij}}{\partial {r}_{ij}}
   \label{continuity-equation-discrete}
\end{equation}
\begin{equation}
    \frac{\text{d} \mathbf v_i^p}{\text{d} t} = -2 \sum_{j} m_j \frac{P^*}{\rho_i \rho_j} {\nabla_i W_{ij}}
   \label{normal-accelaration-discrete}
\end{equation}
Here, $W_{ij}$ represents $W({\mathbf r}_i- {\mathbf r}_j, h)$, which is the kernel function. ${\mathbf r}$ is particle position and ${h}$ is the smoothing length. The subscripts ${i}$ and ${j}$ donate particle numbers, and $m$ is the particle mass. 
$\mathbf e_{ij}$ is the unit vector pointing from particle ${j}$ to particle ${i}$ and $ \mathbf v_{ij}=\mathbf v_i-\mathbf v_j$.
${\nabla_i W_{ij}}=\frac{\partial  W({r}_{ij}, h)}{\partial { r}_{ij}} \mathbf e_{ij}$ is the derivative of kernel function, and ${r}_{ij} = |\mathbf r_{i} - \mathbf r_{j}|$ is the distance between two particles.
$U^*$ and $P^*$, which are obtained from the low-dissipation Riemann solver \cite{zhang2021sphinxsys, zhang2017weakly}, are the solutions of an inter-particle Riemann problem along the unit vector pointing from particle $i$ to $j$. 

Then the shear acceleration $\dot{\mathbf v}^s$ can be discretized by
\begin{equation}
    \frac{\text{d} \mathbf v_i^s}{\text{d} t} = \sum_{j} m_j \frac{\bm{\sigma}^s_i + \bm{\sigma}^s_j}{\rho_i \rho_j} \cdot{\nabla_i W_{ij}}
   \label{shear-accelaration-discrete-nested}
\end{equation}
Refer to Eq. \eqref{shear-stress-integral}-Eq. \eqref{strain-rate}, the velocity gradient needs to be discretized for computing shear stress $\bm{\sigma}^s_i$ and $\bm{\sigma}^s_j$, which has the following form  \cite{espanol2003smoothed}
\begin{equation}
    {\nabla \mathbf v}=\sum_{j} \mathbf v_{ij} {\nabla_i W_{ij}} V_j
   \label{velocity-gradient-discrete}
\end{equation}
where $V_j$ is the volume of particle $j$.

Combined Eq. \eqref{shear-stress-integral}-Eq. \eqref{strain-rate} with Eq. \eqref{shear-acc}, it is found that when calculating shear acceleration $\dot{\mathbf v}^s$ within a time step, we first compute velocity gradient ${\nabla \mathbf v}$ with Eq. \eqref{velocity-gradient-discrete}, then update shear strain rate ${\dot{\bm \varepsilon}}$, followed by computing shear stress rate ${\dot{\bm{\sigma}}^s}$, and subsequently calculate shear stress $\bm{\sigma}^s$. 
Finally, the shear acceleration can be estimated by the divergence of shear stress $\bm{\sigma}^s$ with Eq. \eqref{shear-accelaration-discrete-nested}. 
This nested formulation of shear acceleration will trigger hourglass modes, which is the origin of tensile instability in SPH simulation.
%%%%%%%%%%%%%%%%%%%%%%%%%%%%%%%%%%%%%%%%%%%%%%%%%%%%%%%%%%%%%
%
% 4 Essentially non-hourglass formulation
%
%%%%%%%%%%%%%%%%%%%%%%%%%%%%%%%%%%%%%%%%%%%%%%%%%%%%%%%%%%%%%
\section{Essentially non-hourglass SPH formulation}
\label{non-hourglass-formulation-section}

The formulation in section \ref{original-SPH-formulation} suffers from hourglass issues induced by zero energy modes \cite{vignjevic2000treatment}, which is characterized by a pattern of particle displacement that does not correspond to rigid body motion but still results in zero strain energy \cite{vignjevic2000treatment, vignjevic2009review, swegle1994analysis}.
In a recent effective remedy for hourglass model in TLSPH 
\cite{wu2023essentially},
the particle acceleration due to the divergence of shear stress 
is directly obtained
from a one-step Laplacian formulation of the particle displacement 
other than the nested implementation of the 2nd-order derivatives 
used in the original TLSPH.     
Actually, such non-nested SPH formulation of Laplacian is 
widely used SPH fluid dynamics for computing the viscous-force term
in the Navier-Stokes equations \cite{morris1997modeling, hu2006multi, Monaghan2005SmoothedPH}, 
and is found much stabler than the nested counterpart. Inspired by these previous solutions, we apply a Laplacian operator to calculate the shear acceleration directly in ULSPH simulations of elastic dynamics.

Firstly, we re-formulated the theoretical expression for shear acceleration by combining Eq. \eqref{shear-stress-integral}-Eq. \eqref{shear-acc}

\begin{equation}
    \dot{\mathbf v}^s=\frac{G}{\mathbf \rho}\nabla \cdot {\int_{0}^{t} {(\nabla \mathbf v + (\nabla \mathbf v)^T - \frac{2}{d}\nabla \cdot \mathbf v \mathbf I)}  \text{d}t} 
   \label{nested-formulation}
\end{equation}
By substituting the first Hamiltonian operator on the right-hand side into the integral symbol, the following equation can be obtained
\begin{equation}
    \dot{\mathbf v}^s=\frac{G}{\mathbf \rho} {\int_{0}^{t} {(\nabla^2 \mathbf v + \nabla \nabla \cdot \mathbf v - \frac{2}{d}\nabla \nabla \cdot \mathbf v)}  \text{d}t} 
   \label{non-hourglass-formulation}
\end{equation}
When considering weakly compressible conditions, i.e., ${\nabla \cdot \mathbf v \approx}$ 0, Eq. \eqref{non-hourglass-formulation} can be simplified to
\begin{equation}
    {\dot{\mathbf v}^s} \approx {\frac{G}{\mathbf \rho} {\int_{0}^{t} {(\nabla^2 \mathbf v)}  \text{d}t}} 
   \label{simplify-non-hourglass-formulation}
\end{equation}
The Laplacian operator needs to be discretized for calculating the shear acceleration. Refer to the literature \cite{espanol2003smoothed}, $\nabla^2 \mathbf v$ can be discretized as
\begin{equation}
    {\nabla^2 \mathbf v} = 2 \sum_{j} \frac{\mathbf v_{ij}}{r_{ij}} \frac{\partial  W_{ij}}{\partial {r}_{ij}} V_j
   \label{2nd-derivative-discrete}
\end{equation}
However, the aforementioned formula does not fulfill the requirement for angular-momentum conservation. 
In other words, using this discretized form of the second derivative of velocity  to compute shear acceleration in Eq.  \eqref{simplify-non-hourglass-formulation} cannot remove the effects of rigid rotation. 
Based on the research of Hu et al. \cite{hu2006angular}, we consider an angular-momentum conservative form for the second derivative of velocity, to eliminate the contribution of rigid rotation to the shear acceleration that should not have occurred in the first place, as shown in Eq.  \eqref{2nd-derivative-discrete-angular}. 
\begin{equation}
    {\nabla^2 \mathbf v} = 2 \zeta {\sum_{j} \frac{\mathbf e_{ij} \cdot \mathbf v_{ij}}{r_{ij}} {\nabla_i W_{ij}} V_j}
   \label{2nd-derivative-discrete-angular}
\end{equation}
where $\zeta$ is a parameter related to the smoothing length $h$ and the type of kernel function. $\zeta$ needs to be calibrated with numerical experiments. Then the non-nested formulation of shear acceleration can be written as
\begin{equation}
    {\dot{\mathbf v}^s} = 2 \zeta {\frac{G}{\mathbf \rho} {\int_{0}^{t} {\left(\sum_{j} \frac{\mathbf e_{ij} \cdot \mathbf v_{ij}}{r_{ij}} {\nabla_i W_{ij}} V_j \right)}  \text{d}t}} 
   \label{non-hourglass-formulation-discrete-angular}
\end{equation}

This is the final form of shear acceleration without hourglass modes, and thus can eliminate the tensile instability in essence. 
In Section \ref{2D-oscillating-plate}, we will provide a demonstration of the disparities between angular-momentum conservative and non-conservative approaches in numerical computations.
It should be noted that the shear acceleration calculated at time step ${n}$ will be used at the next time step ${n+1}$. 

%%%%%%%%%%%%%%%%%%%%%%%%%%%%%%%%%%%%%%%%%%%%%%%%%%%%%%%%%%%%%
%
% 5 Dual-criteria time stepping 
%
%%%%%%%%%%%%%%%%%%%%%%%%%%%%%%%%%%%%%%%%%%%%%%%%%%%%%%%%%%%%%
\section{Dual-criteria time stepping}\label{time-integration}

As we mentioned in Section \ref{introduction}, due to the necessity of updating particle configurations at each computational time step, a persistently challenging issue in ULSPH is its low computational efficiency.
In this section, the dual-criteria time stepping originally proposed for fluid simulations \cite{zhang2020dual} is introduced to solid simulations for the first time, to improve the calculation efficiency by reducing the frequency for updating particle configurations, while maintain high computation accuracy at the same time. 

The dual-criteria time stepping strategy employs a larger advection time step $\bigtriangleup t_{ad}$, and a smaller acoustic time step $\bigtriangleup t_{ac}$. The particle configuration is updated in the advection time step $\bigtriangleup t_{ad}$, which is defined as

\begin{equation}
    \bigtriangleup t_{ad} = CFL_{ad}\frac{h}{\left\lvert \mathbf v \right\rvert_{max} }
   \label{advection-time-step}
\end{equation}
where $CFL_{ad}=0.2$, $\left\lvert \mathbf v \right\rvert_{max}$ is the maximum particle advection speed and $h$ is the smoothing length. The acoustic time step $\bigtriangleup t_{ac}$, involving the update of particle properties such velocity and density, has the following form
\begin{equation}
    \bigtriangleup t_{ac} = CFL_{ac}\frac{h}{c_0+\left\lvert \mathbf v \right\rvert_{max} }
   \label{acoustic-time-step}
\end{equation}
where $CFL_{ac}=0.4$ and $c_0$ is the sound speed. 

Then the position-based Verlet scheme is applied for the acoustic time integration \cite{zhang2021multi}. 
The beginning of the acoustic time step is indicated by superscript $n$, and the midpoint and new time step are donated by superscript $n+\frac{1}{2}$ and $n+1$ respectively. In the Verlet scheme, the particle position and density are firstly updated to the midpoint with
\begin{equation}
	\begin{cases}
		{\mathbf r}^{n+\frac{1}{2}}={\mathbf r}^n+ \frac{1}{2} {\bigtriangleup t_{ac}} {\mathbf v}^n  \\
		{\rho}^{n+\frac{1}{2}}={\rho}^n+ \frac{1}{2} {\bigtriangleup t_{ac}} ({\frac{\text{d} \rho}{\text{d} t}})^n        \\
	\end{cases}
   \label{time-step-half}
\end{equation}
Then the velocity is updated to the new time step after the particle acceleration is determined.
\begin{equation}
    {\mathbf v}_{n+1}={\mathbf v}_{n} + {\bigtriangleup t_{ac}} ({\frac{\text{d} \mathbf v}{\text{d} t}})^n
   \label{time-step-velocity}
\end{equation}
Finally, the particle position and density are updated to the new time step by
\begin{equation}
	\begin{cases}
		{\mathbf r}^{n+1}={\mathbf r}^{n+\frac{1}{2}}+ \frac{1}{2} {\bigtriangleup t_{ac}} {\mathbf v}^{n+1}  \\
		{\rho}^{n+1}={\rho}^{n+\frac{1}{2}}+ \frac{1}{2} {\bigtriangleup t_{ac}} ({\frac{\text{d} \rho}{\text{d} t}})^{n+1}        \\
	\end{cases}
   \label{time-step-full}
\end{equation}

%%%%%%%%%%%%%%%%%%%%%%%%%%%%%%%%%%%%%%%%%%%%%%%%%%%%%%%%%%%%%
%
% 6 Numerical examples
%
%%%%%%%%%%%%%%%%%%%%%%%%%%%%%%%%%%%%%%%%%%%%%%%%%%%%%%%%%%%%%
\section{Numerical examples}\label{validation}

In this section, several benchmark cases are tested, and the results are compared with analytical solutions or the results from other numerical studies, from both qualitative and quantitative perspectives. 
Specifically,  our results are compared with the results obtained using original nested formulation and nested formulation with artificial stress \cite{gray2001sph}, to illustrate the calculation accuracy, stability, and robustness of the present method. 
Several abbreviations for different SPH methods are defined to facilitate the writing and reading of the article. 
The original SPH method is defined as "SPH-OG"; the original SPH method with introducing artificial stress \cite{gray2001sph} is represented by "SPH-OAS"; 
the present essentially non-hourglass formulation is expressed as "SPH-ENOG" in the following text. 
There are two artificial coefficients in the artificial stress term, and the selection of them refers to the literature \cite{gray2001sph} in this section.

The 5th-order Wendland kernel \cite{wendland1995piecewise} with a smoothing length of $h$ = 1.3$dp$ and the cut-off radius equals to 2.6$dp$, where $dp$ is the initial particle spacing, is applied for all the cases in this paper. 
Based on the selected kernel type and smoothing length, the coefficient $\zeta$ in Eq. \eqref{non-hourglass-formulation-discrete-angular} is set to $0.7d$+2.1 (3.5 for 2D situations and 4.2 for 3D situations) according to numerical experiments and is adopted throughout this study. 
All the physical quantities in this article are presented in dimensionless form. 

In this newly-developed non-hourglass formulation, we skip the calculation of shear stress and the shear acceleration is computed directly. The shear stress can be obtained separately by Eq. \eqref{shear-stress-integral}-Eq. \eqref{strain-rate} if needed. The velocity gradient in Eq. \eqref{strain-rate} is discretized by
\begin{equation}
    {\nabla \mathbf v}=\sum_{j} \mathbf v_{ij}  (\mathbf B_i {\nabla_i W_{ij}}) V_j
   \label{velocity-gradient-discrete-correction}
\end{equation}
where $\mathbf B_i$ is the correction matrix for kernel gradient \cite{randles1996smoothed, bonet2002simplified, zhang2021integrative} and is defined as
\begin{equation}
    \mathbf B_i = \left({\sum_{j} \mathbf r_{ij} {\nabla_i W_{ij}}  V_j} \right)^{-1}
   \label{correction-matrix}
\end{equation}

All the simulations in this section are run on a CentOS-8 system with 32 cores. The detail information of the CPU is "64 Intel(R) Xeon(R) Gold 6226R CPU @ 2.90GHz". 
%%%%%%%%%%%%%%%%%%%%%%%%%%%%%%%%%%%%%%%%%%%%%%%%%%%%%%%%%%%%%
% 6.1 2D oscillating plate
%%%%%%%%%%%%%%%%%%%%%%%%%%%%%%%%%%%%%%%%%%%%%%%%%%%%%%%%%%%%%
\subsection{2D oscillating plate}
\label{2D-oscillating-plate}

\begin{figure}[htb!]
	\centering
	\includegraphics[trim = 0cm 0cm 0cm 0cm, clip,width=.85\textwidth]{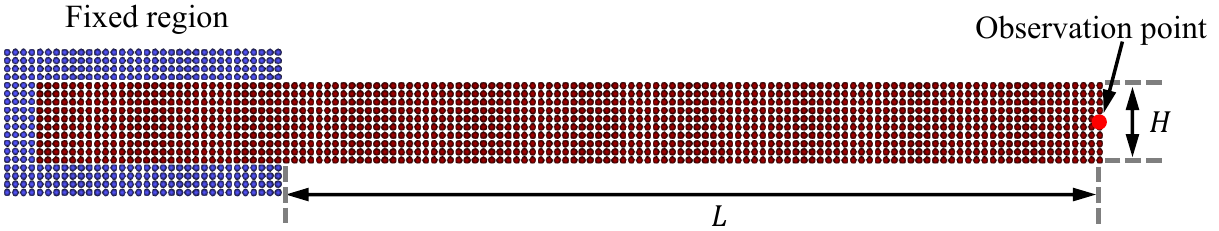}
	\caption{Model setup for 2D oscillating plate}
	\label{figs:2D-plate-setup}
\end{figure}

As shown in Fig. \ref{figs:2D-plate-setup}, a 2D plate with one edge fixed is firstly used to verify the proposed method, and the results are compared with previous theoretical \cite{landau2013course} and numerical \cite{gray2001sph} solutions. The length and thickness of the plate are $L$ and $H$ respectively, and the left part is fixed to produce a cantilever plate. An observation point is set at the middle of the tail, and the vertical displacement of the point is recorded as the amplitude. An initial velocity $v_y$, which perpendicular to the plate strip, is applied with

\begin{equation}
    v_y(x) = v_f c_0 \frac{f(x)}{f(L)}
   \label{2D-plate-initial-velocity}
\end{equation}
where the constant $v_f$ is an input parameter and

\begin{equation}
    \begin{aligned}
        f(x)=(\sin (kL) + \sinh(kL))(\cos(kx)-\cosh(kx)) \\
        - (\cos(kL)+\cosh(kL))(\sin(kx)-\sinh(kx))
    \end{aligned}
   \label{2D-plate-fx}
\end{equation}
where $kL=1.875$ is determined by $\cos(kL) \cosh(kL)=-1$. The frequency $\omega $ of the oscillating plate is theoretically given by
\begin{equation}
    {\omega}^2=\frac{EH^2k^4}{12\rho_0 (1-{\nu}^4 )}
   \label{2D-plate-frequency}
\end{equation}

The material and dimensional parameters in this case follow literatures \cite{gray2001sph, zhang2017generalized}, i.e., density $\rho_0=1000$, Young's modulus $E=2\times 10^6$, Poisson's ratio $\nu = 0.3975$, $L=0.2$, and $H=0.02$.

\begin{figure}[htb!]
	\centering
	\includegraphics[trim = 0cm 0cm 0cm 0cm, clip,width=1.0\textwidth]{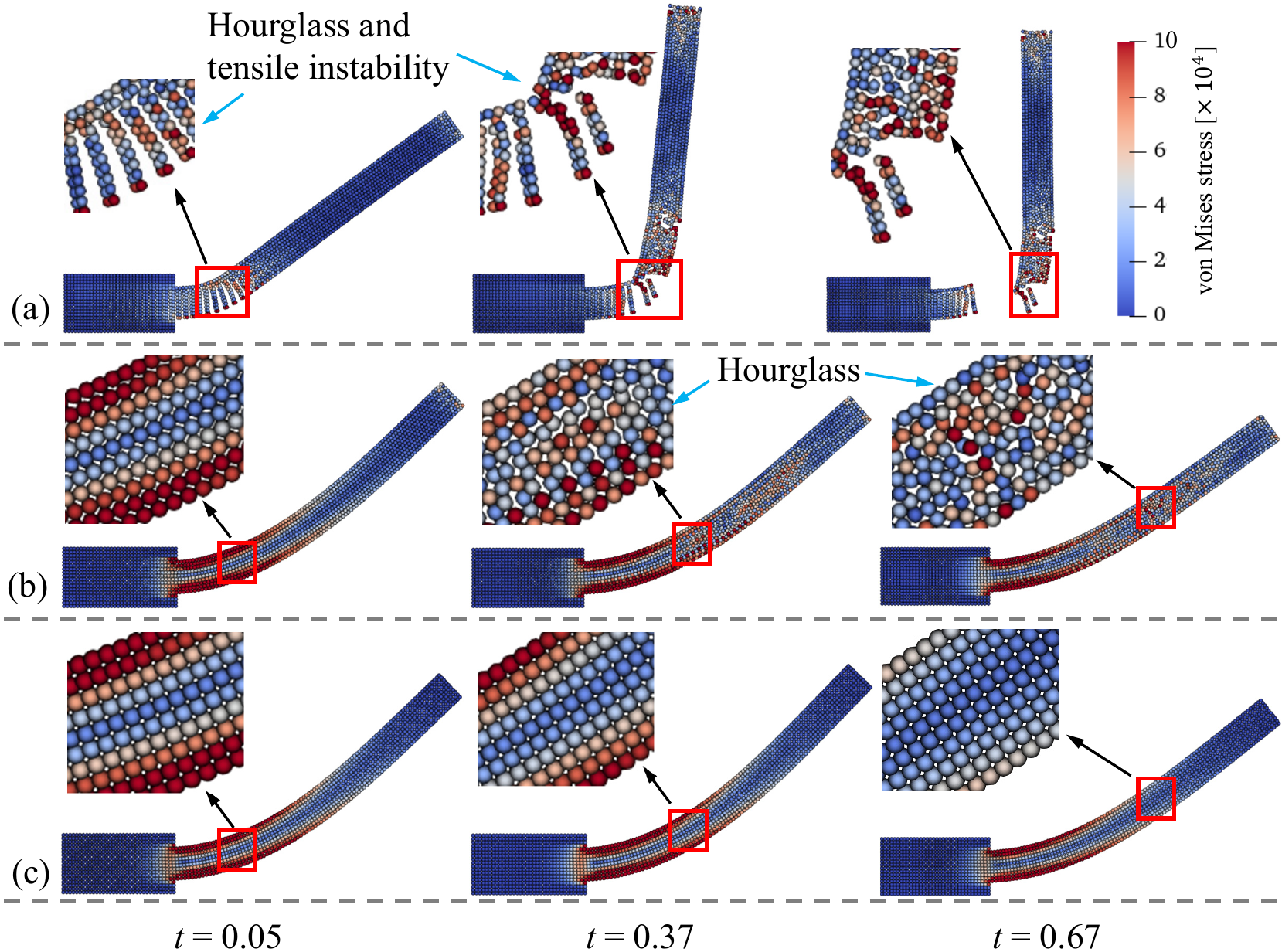}
	\caption{Evolution of particle configuration with time (t=0.05, 0.37 and 0.67) for (a) SPH-OG, (b) SPH-OAS, and (c) SPH-ENOG. The particles are colored by von Mises stress.}
	\label{figs:2D-plate-nested-non-hourglass}
\end{figure}

As shown in Fig. \ref{figs:2D-plate-nested-non-hourglass}(a), SPH-OG leads to hourglass and tensile instability issues when simulating elastic deformation. 
Numerical fractures occur at the beginning of the simulation ($t=0.05$), which is the tensile instability addressed by previous researchers \cite{gray2001sph, swegle1995smoothed}; 
the nonphysical zigzag particle distribution and the non-uniform profile of von Mises stress indicate the hourglass mode. 
Fig. \ref{figs:2D-plate-nested-non-hourglass}(b) shows the results obtained by SPH-OAS, in which the tensile instability can be suppressed. However, the hourglass still occurs and became visually evident when $t=0.37$. This is because the error in the nested formulation is in integral form, which gradually accumulates over time. 
The results produced by the present SPH-ENOG are shown in Fig. \ref{figs:2D-plate-nested-non-hourglass}(c). Clearly, neither hourglass nor tensile instability appears even when the time $t=0.67$. The particle distribution is still uniform, and the stress profile is smooth.

\begin{figure}[htb!]
	\centering
	\includegraphics[trim = 0cm 0cm 0cm 0cm, clip,width=.65\textwidth]{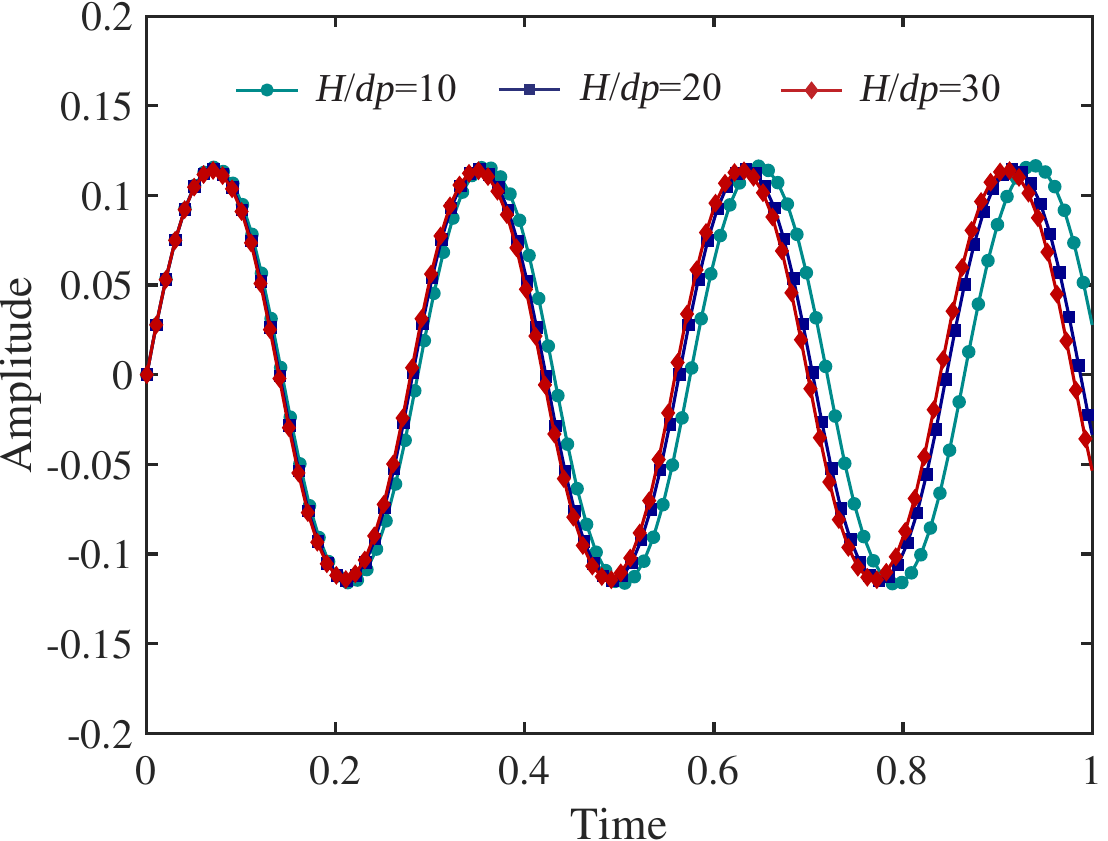}
	\caption{Convergence verification for the 2D oscillating plate with the present SPH-ENOG. The curves show the changes of amplitude with time. Here, $L$=0.2, $H$=0.02, and $\mathbf v_f$=0.05.}
	\label{figs:2D-plate-resolutions}
\end{figure}

The convergence of the present new formulation is validated, as shown in Fig. \ref{figs:2D-plate-resolutions}. Three cases with different resolutions ($H/dp=10$, $H/dp=20$ and $H/dp=30$) are tested and the variation of amplitudes over time are illustrated. It can be observed that, with the increase of resolution, the differences between different solutions are deceasing, which is consistent with the results in literatures \cite{gray2001sph, zhang2017generalized, wu2023essentially} and indicates the convergence of this present algorithm. 

\begin{figure}[htb!]
	\centering
	\includegraphics[trim = 0cm 0cm 0cm 0cm, clip,width=0.9\textwidth]{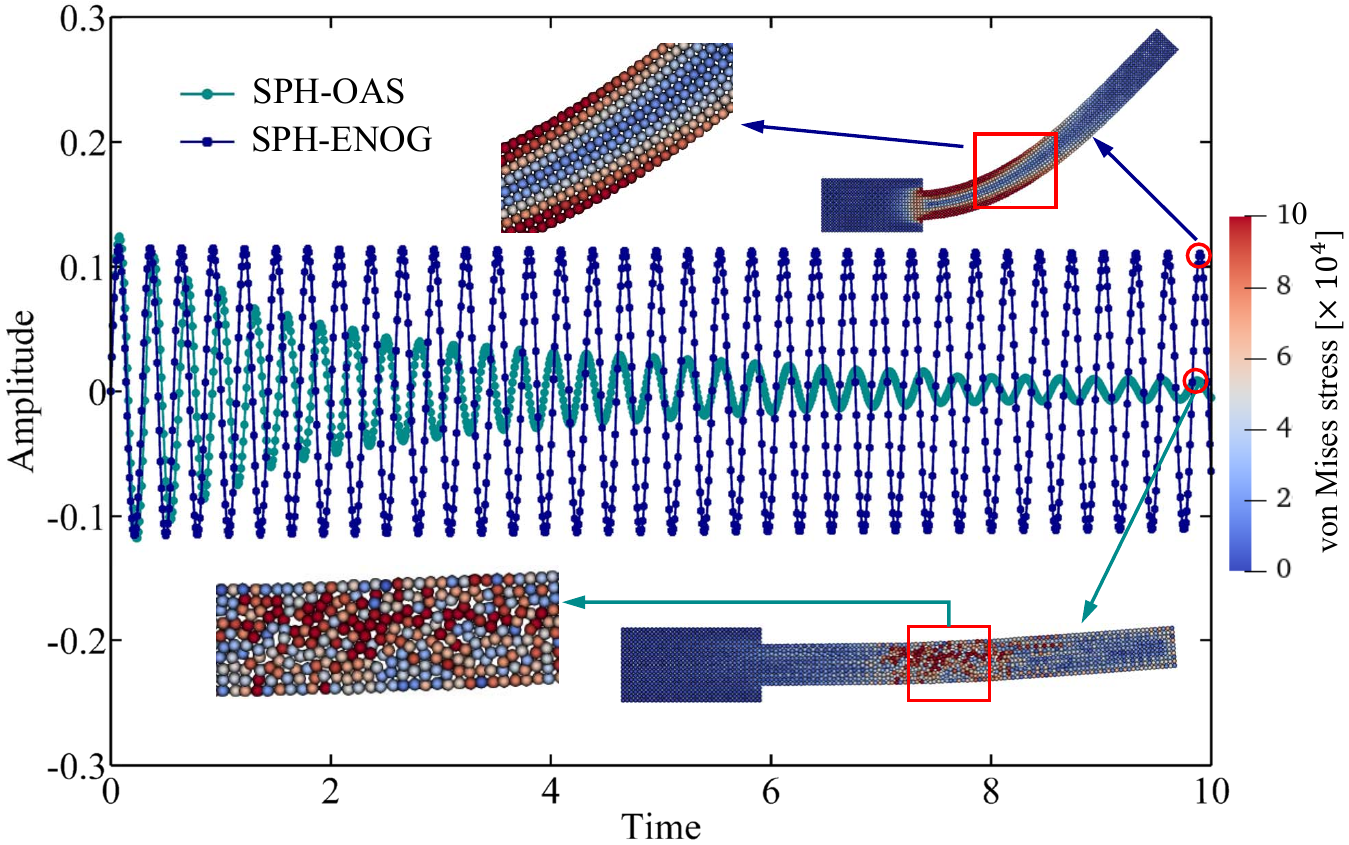}
	\caption{Test the long-term stability for the SPH-ENOG, the result is compared with the SPH-OAS. Here, $L$=0.2, $H$=0.02, $H/dp=10$ and $\mathbf v_f$=0.05. The particles are colored by von Mises stress.}
	\label{figs:2D-plate-stress-test}
\end{figure}

Then a stress testing is performed with a long-time simulation, to check the stability of the current algorithm. 
As shown in Fig. \ref{figs:2D-plate-stress-test}, the simulation lasts for over 30 oscillations, and the result from SPH-OAS is also illustrated for comparison. 
Here, the two simulations are run with single time step \cite{zhang2017generalized}, as we want to minimize the accumulated integration error in long-time simulations. 
The images above and below the curves in Fig. \ref{figs:2D-plate-stress-test} respectively represent the particle distribution obtained using SPH-ENOG and SPH-OAS at a time around 10. 
As depicted in Fig. \ref{figs:2D-plate-stress-test}, with the proposed SPH-ENOG, the particle and stress distribution still keep uniform until the end of the simulation; while for SPH-OAS, the hourglass issue appears to be serious at $t\approx 10$. 
Moreover, with the present SPH-ENOG, the amplitude only decrease marginally at $t$ around 10 compared with the amplitude at $t=0$. The slight decrease in amplitude over time is due to the numerical dissipation introduced in the Riemann solver \cite{zhang2017weakly}.
On the contrary, the SPH-OAS exhibits rapid energy decay, thus it cannot be used for long-duration computations.

Furthermore, the accuracy is checked and verified with theoretical solutions and SPH-OAS from Gray et al.'s work \cite{gray2001sph}. As shown in Table \ref{2D-plate-error}, the first period of oscillation is recorded for both SPH-ENOG and SPH-OAS with $L$=0.2, $H$=0.02 and $H/dp=30$. 
Compared with the analytical solution, the errors of the SPH-ENOG are at the same level as SPH-OAG, confirming the accuracy of the current SPH-ENOG. It should be noted that the analytical solution are obtained based on a thin plate model. If the thickness of the plate is reduced to $H$=0.01, the error with $v_f$=0.001 for the SPH-ENOG also decrease to around 0.6$\%$, which is in consistent with previous studies \cite{gray2001sph, zhang2017generalized}.

\begin{table}
	\scriptsize
	\centering
	\caption{Comparison of oscillation periods $T$ obtained from the present SPH-ENOG, SPH-OAS and analytical solutions. Here, $L$=0.2, $H$=0.02 and $H/dp=30$.}
	\begin{tabularx}{8.5cm}{@{\extracolsep{\fill}}lcccc}
		\hline
		$v_f$ & 0.001 & 0.01 & 0.03 & 0.05\\
		\hline
        $T$ (Analytical) & \multicolumn{4}{c}{0.254}   \\
		$T$ (SPH-ENOG) & 0.262 & 0.263 & 0.268 & 0.279 \\
		$T$ (SPH-OAS) & 0.273 & 0.273 & 0.275 & 0.278 \\
        Error (SPH-ENOG) & 3.1$\%$ & 3.5$\%$ & 5.5$\%$ & 9.8$\%$ \\
		Error (SPH-OAS) & 7.5$\%$ & 7.5$\%$ & 8.3$\%$ & 9.4$\%$ \\
		\hline
	\end{tabularx}
	\label{2D-plate-error}
\end{table}

\begin{figure}[htb!]
	\centering
	\includegraphics[trim = 0cm 0cm 0cm 0cm, clip,width=0.75\textwidth]{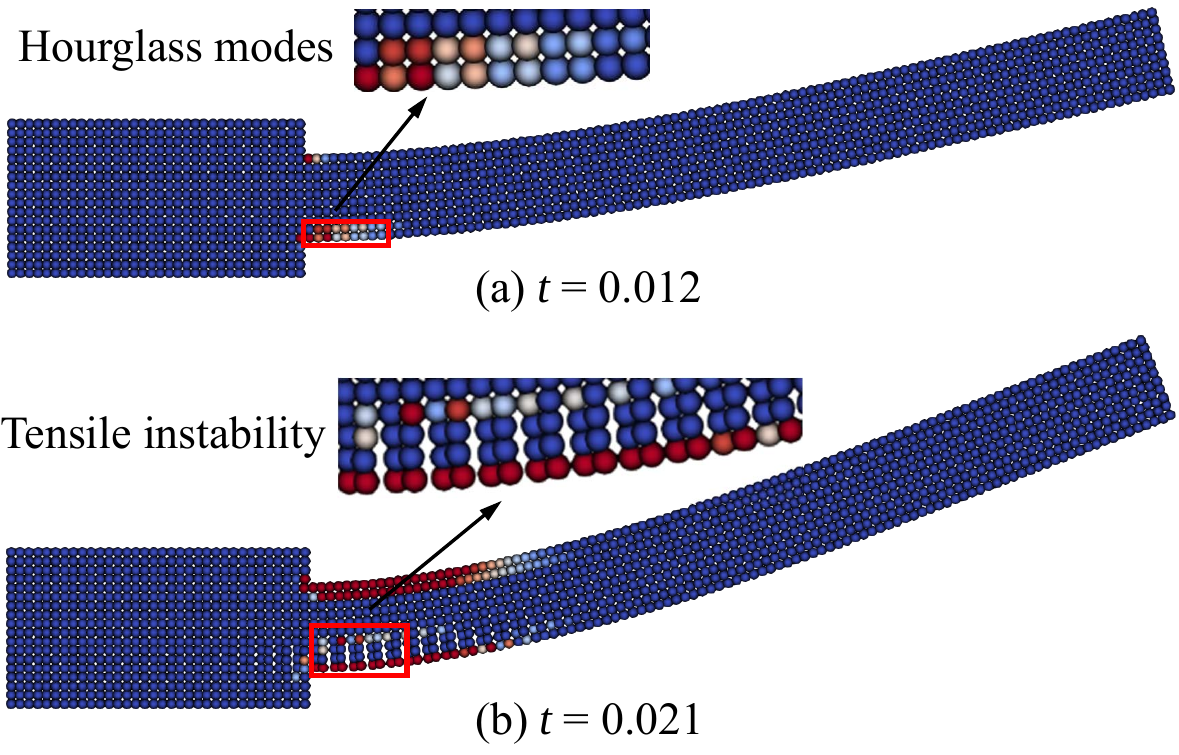}
	\caption{Illustration for occurrences of hourglass modes and tension instability in 2D oscillating plates with the SPH-OG. The hourglass issue appears at the initial stages when (a) $t=0.012$, and the tensile instability can be observed later at (b) $t=0.021$. The particles are colored by von Mises stress.}
	\label{figs:tensile-instability-after-hourglass}
\end{figure}

Simultaneously, we provided visual evidence to demonstrate that the tensile instability occurs after hourglass modes. As shown in Fig. \ref{figs:tensile-instability-after-hourglass}, when we simulate the 2D oscillating plate using the SPH-OG, we first observe the appearance of hourglass modes at the initial stage ($t=0.012$), and tension instability gradually emerges afterwards at $t=0.021$.

\begin{figure}[htb!]
	\centering
	\includegraphics[trim = 0cm 0cm 0cm 0cm, clip,width=0.95\textwidth]{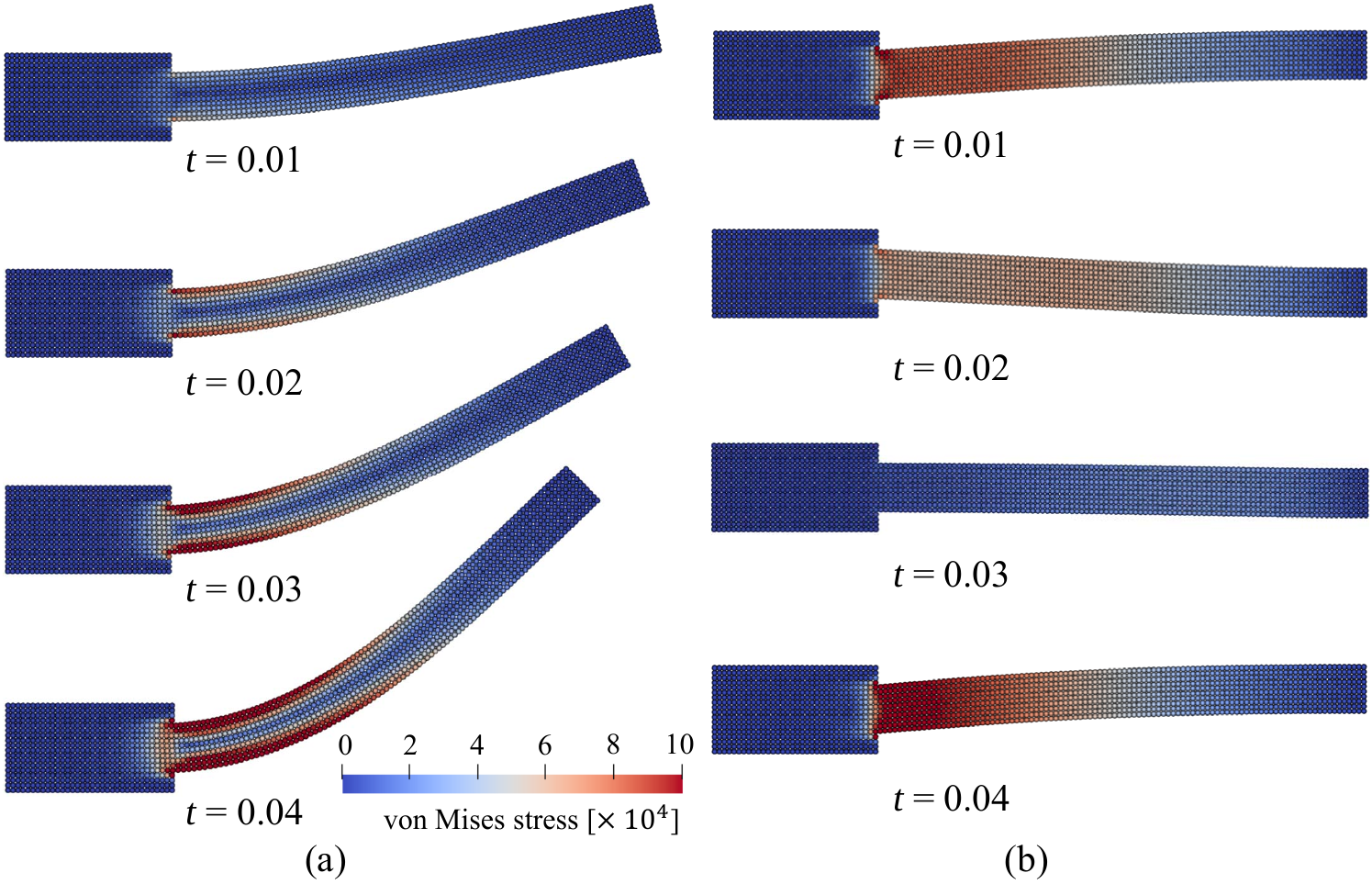}
	\caption{Demonstration of the differences in the behavior of oscillating plates when using angular-momentum conservative and non-conservative forms to discretize the second derivative of velocity gradient: (a) Angular-momentum conservative form; (b) Angular-momentum non-conservative form. Here, $L$=0.2, $H$=0.02, $H/dp=10$ and $\mathbf v_f$=0.05. The particles are colored by von Mises stress.}
	\label{figs:angular-momentum-conservative}
\end{figure}

Next, we present the effects of employing angular-momentum conservation (Eq. \eqref{2nd-derivative-discrete-angular}) and non-conservation (Eq. \eqref{2nd-derivative-discrete}) approaches in discretizing the second derivative of velocity gradient and their implications on the results. As shown in Fig. \ref{figs:angular-momentum-conservative}, it can be observed that when applying the non-conservative form, the behavior of the oscillating plate does not oscillate as expected, but exhibits a strong resistance to motion. This is because, the influence of rigid rotation is not eliminated in the non-conservative form, which means that rigid rotation can also produce significant shear forces, thereby hindering the plate's motion. Correspondingly, the behavior of the plate can be correctly estimated with the angular-momentum conservative form, demonstrating the necessity of adopting this conservative type.

The performance of dual-criteria time stepping scheme is tested with the SPH-ENOG, and the results from previous single time stepping method \cite{zhang2017generalized} is used for comparison. As shown in Table \ref{2D-plate-efficiency}, the simulation ends at physical time $t=1$, and the wall-clock time spent for dual-criteria (donate as $T_d$) and single-criteria (donate as $T_s$) time stepping is recorded. 
It can be seen the computing efficiency of the dual-criteria time steeping is approximately twice that of the previous method.

\begin{table}
	\scriptsize
	\centering
	\caption{Comparison of computational efficiency for dual-criteria ($T_d$) and single-criteria ($T_s$) time stepping scheme with 2D oscillating beams. Here, $L$=0.2, $H$=0.02 and $\mathbf v_f$=0.05. The simulation stops at physical time $t=1$ and the wall-clock time spent is recorded. $N_p$ represents the total particle number.}
	\begin{tabularx}{8.5cm}{@{\extracolsep{\fill}}lcccc}
		\hline
		$H/dp$ & 10 & 20 & 30 \\
		\hline
		$N_p$ (k) & 1.6 & 5.8 & 12.6 \\
		$T_d$ & 25.8 & 66.3 & 130.1 \\
		$T_s$ & 43.0 & 139.5 & 333.0 \\
		\hline
	\end{tabularx}
	\label{2D-plate-efficiency}
\end{table}

%%%%%%%%%%%%%%%%%%%%%%%%%%%%%%%%%%%%%%%%%%%%%%%%%%%%%%%%%%%%%
% 6.2 3D oscillating plate
%%%%%%%%%%%%%%%%%%%%%%%%%%%%%%%%%%%%%%%%%%%%%%%%%%%%%%%%%%%%%
\subsection{3D oscillating plate}
\label{3D-oscillating-plate}

\begin{figure}[htb!]
	\centering
	\includegraphics[trim = 0cm 0cm 0cm 0cm, clip,width=0.6\textwidth]{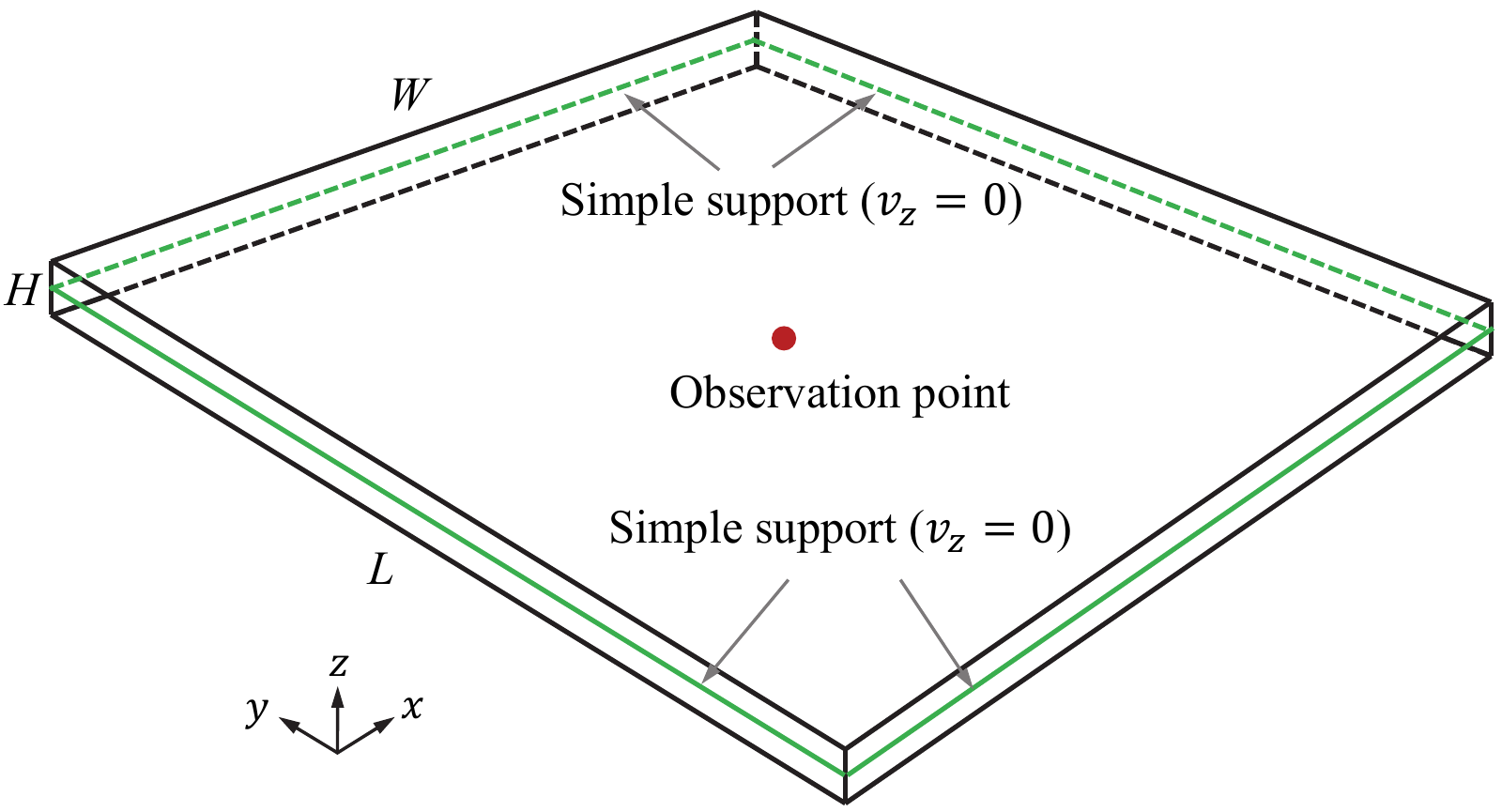}
	\caption{Model setup for the 3D oscillating plate, and the observation point locates at the center of the plate.}
	\label{figs:3D-plate-setup}
\end{figure}

The oscillation of a 3D thin plate, with a simple support boundary condition for all lateral edges,  is considered in this section. 
As shown in Fig. \ref{figs:3D-plate-setup}, a square plate with length $L=0.4$, width $=0.4$ and height $H = 0.01$ is constructed \cite{khayyer20213d,wu2023essentially,leissa1969vibration}.
A simple boundary condition is applied to the particles in the middle of the four lateral sides. Specifically, the displacement of these particles along the $z$-direction is fixed. An observation point is set at the center of the plate.
The particles are subjected to the initial velocity $v_z$
\begin{equation}
    {v_z}(x,y)=\sin{\frac{m \pi x}{L}} \sin{\frac{n \pi y}{W}}
   \label{3D-plate-vy}
\end{equation}
where $m$ and $n$ donate integers controlling the vibration mode in $x$ and $y$ directions respectively. 
The theoretical vibration period for the 3D thin plate is given by
\begin{equation}
    T=\frac{2}{\pi} \left[\left({\frac{m}{L}}\right)^2+\left({\frac{n}{W}}\right)^2 \right] ^{-1} {\sqrt{\frac{\rho_0 H}{D}}}
   \label{3D-plate-period}
\end{equation}
where $D$ represents the flexural rigidity and is defined as
\begin{equation}
    D=\frac{EH^3}{12(1-\nu^2)} 
   \label{3D-plate-parameter-D}
\end{equation}
The material parameters are set as follows: density $\rho_0=1000$, Young's modulus $E=1\times 10^8$, and Poisson's ratio $\nu = 0.3$.

\begin{figure}[htb!]
	\centering
	\includegraphics[trim = 0cm 0cm 0cm 0cm, clip,width=0.9\textwidth]{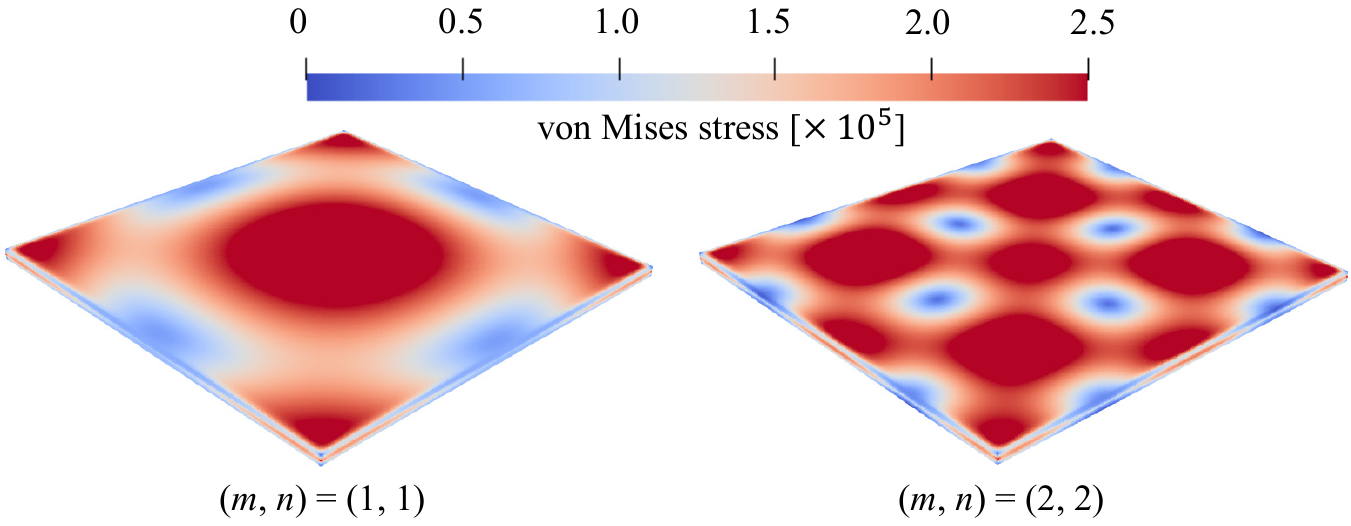}
	\caption{3D oscillating plates obtained by the SPH-ENOG at time $t=0.01$ for vibration modes $(m,n)=(1,1)$ and $(2,2)$. The figures are colored by von Mises stress and the spatial particle discretization $H/dp = 9$.}
	\label{figs:3D-plate-stress-profile}
\end{figure}

\begin{figure}[htb!]
	\centering
	\includegraphics[trim = 0cm 0cm 0cm 0cm, clip,width=0.7\textwidth]{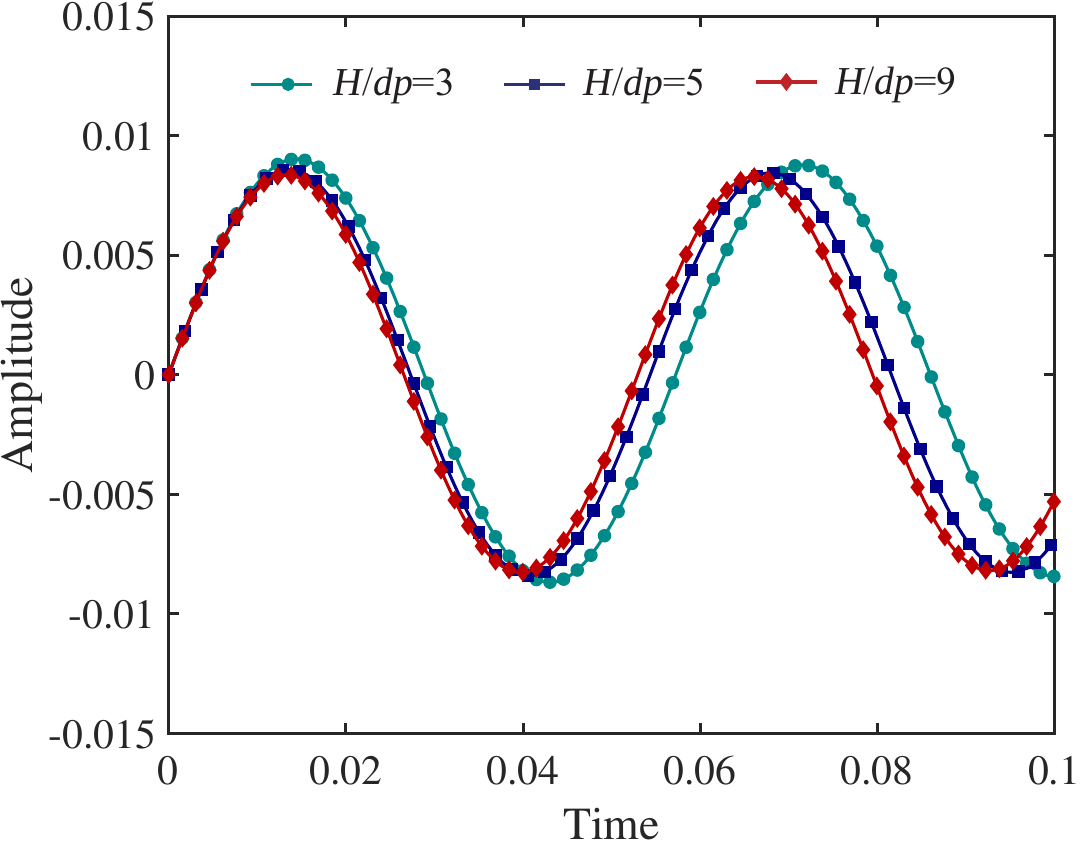}
	\caption{Convergence verification for the 3D oscillating plate with the present SPH-ENOG. The curves show the changes of amplitude with time. Here, $(m,n)=(1,1)$.}
	\label{figs:3D-plate-convergence}
\end{figure}

Firstly, the particle distribution and the profile of von Mises stress are checked. Fig. \ref{figs:3D-plate-stress-profile} illustrates the deformed particle configuration with von Mises stress contour obtained by the SPH-ENOG at time $t=0.01$ for vibration modes $(m,n)=(1,1)$ and $(2,2)$. The SPH-ENOG can generate a smooth stress profile without hourglass modes and tensile instability.
Fig. \ref{figs:3D-plate-convergence} shows the evolution of the amplitude (displacement along $z$-axis of the observation point) with time for plates with different resolutions. As can be seen, with the increase of resolution (decrease of initial particle spacing $dp$), the period and amplitude of the curve gradually approach a certain value, indicating the convergence of the SPH-ENOG. Moreover, the first periods of oscillations with different $(m,n)$ values and resolutions are calculated and compared with analytical solutions, to further validate the convergence and accuracy of the present SPH-ENOG. As shown in Table \ref{3D-plate-error}, the periods converge rapidly with increasing resolutions and agree well with the analytical solutions.

\begin{table}
	\scriptsize
	\centering
	\caption{Comparison of the first oscillation period $T$ obtained the present SPH-ENOG and analytical solutions.}
	\begin{tabularx}{9.5cm}{@{\extracolsep{\fill}}lcccc}
		\hline
		SPH-ENOG & $H/dp=3$ & $H/dp=5$ & $H/dp=9$ & Analytical\\
		\hline
        $(m,n)=(1,1)$ &  0.0572 & 0.0543  & 0.0529  &  0.0532 \\
		$(m,n)=(2,1)$ & 0.0231 & 0.0218 & 0.0212 & 0.0213 \\
		$(m,n)=(2,2)$ & 0.0148 & 0.0140 & 0.0136 & 0.0133 \\
		\hline
	\end{tabularx}
	\label{3D-plate-error}
\end{table}

The performance of dual-criteria time stepping scheme is tested with the SPH-ENOG for 3D oscillating plates, and the results from previous single time stepping method is used for comparison. As shown in Table \ref{3D-plate-efficiency}, the simulation ends at physical time $t=0.1$, and the wall-clock time spent for dual-criteria (donate as $T_d$) and single-criteria (donate as $T_s$) time stepping is recorded. 
Clearly, when using the dual-criteria time stepping approach, the computational time reduces to approximately one-third of the time required for single-criteria time stepping.

\begin{table}
	\scriptsize
	\centering
	\caption{Comparison of computational efficiency for dual-criteria ($T_d$) and single-criteria ($T_s$) time stepping scheme with 3D oscillating beam. Here, $(m,n)=(2,2)$. The simulation stops at physical time $t=0.1$ and the wall-clock time spent is recorded. $N_p$ represents the total particle number.}
	\begin{tabularx}{8.5cm}{@{\extracolsep{\fill}}lcccc}
		\hline
		$H/dp$ & 3 & 5 & 9 \\
		\hline
		$N_p$ (k) & 44.7 & 204.0 & 1179.4 \\
		$T_d$ & 122.6 & 981.3 & 10629.4 \\
		$T_s$ & 331.2 & 2683.1 & 30273.3 \\
		\hline
	\end{tabularx}
	\label{3D-plate-efficiency}
\end{table}
%%%%%%%%%%%%%%%%%%%%%%%%%%%%%%%%%%%%%%%%%%%%%%%%%%%%%%%%%%%%%
% 6.3 2D colliding rubber rings
%%%%%%%%%%%%%%%%%%%%%%%%%%%%%%%%%%%%%%%%%%%%%%%%%%%%%%%%%%%%%
\subsection{2D colliding rubber rings}\label{rubber-rings}

\begin{figure}[htb!]
	\centering
	\includegraphics[trim = 0cm 0cm 0cm 0cm, clip,width=0.7\textwidth]{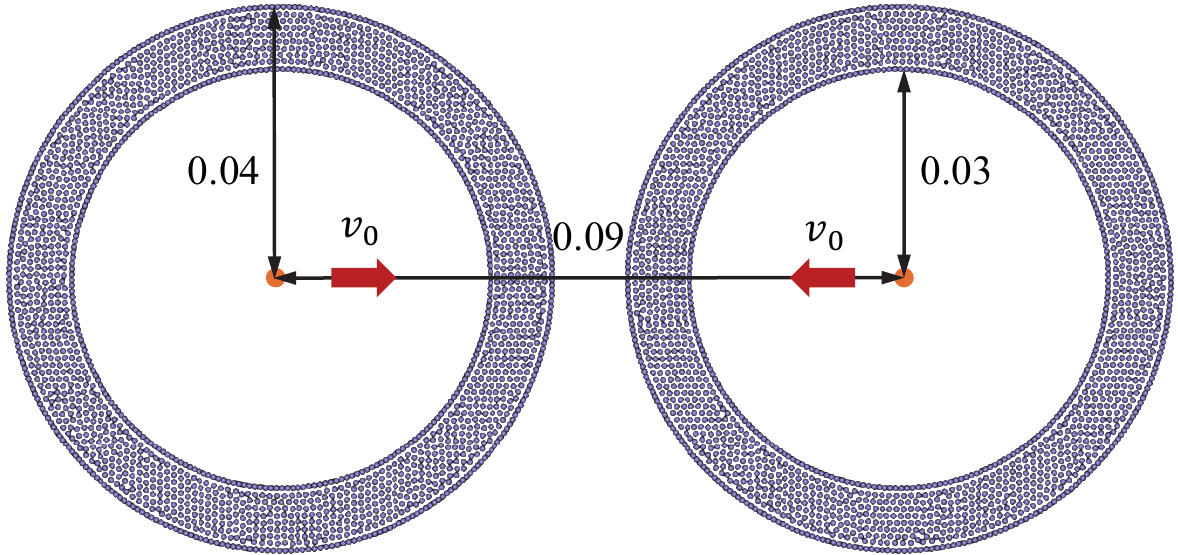}
	\caption{Model setup for 2D colliding rubber rings.}
	\label{figs:2D-ring-setup}
\end{figure}

The collision of two rubber rings is simulated in this section refer to literatures \cite{gray2001sph, monaghan2000sph, zhang2017generalized}. 
As shown in Fig. \ref{figs:2D-ring-setup}, two rings with inner radius 0.03 and outer radius 0.04 are moving towards each other with the initial velocity magnitude $v_0$ (the relative velocity of the two rings is 2$v_0$), and the distance between the centers of the two rings is 0.09. The initial uniform particle distribution is achieved by a level-set based pre-processing technique \cite{yu2023level}. 
When two rings collide with each other, a significant tensile force will be generated. In this case, we will show that the numerical fracture (i.e., tensile instability) produced by SPH-OG and the zigzag particle/stress distribution (i.e., hourglass modes) produced by SPH-OG and SPH-OAS do not occur in the present SPH-ENOG. 
The material parameters are set as follows: density $\rho_0=1200$, Young's modulus $E=1\times 10^7$, and Poisson's ratio $\nu = 0.4$. The initial particle spacing is 0.001.

\begin{figure}[htb!]
	\centering
	\includegraphics[trim = 0cm 0cm 0cm 0cm, clip,width=0.95\textwidth]{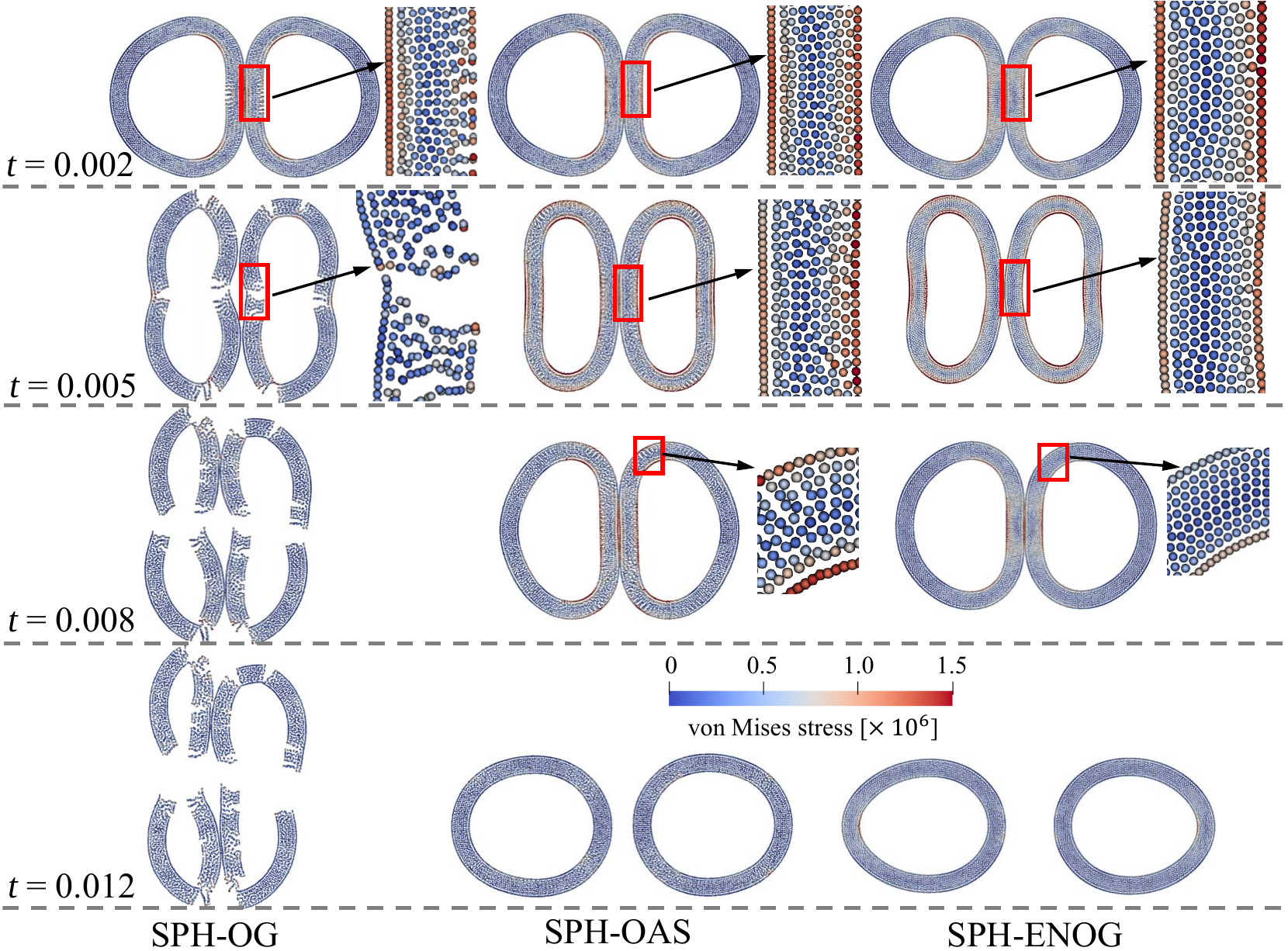}
	\caption{Evolution of particle configuration with time ($t=0.002$, 0.005, 0.008 and 0.012) for 2D colliding rubber rings. The results are obtained by different SPH methods, i.e., SPH-OG (left column), SPH-OAS (middle column), and SPH-ENOG (right column). The initial velocity magnitude $v_0=0.06c_0$ and the figures are colored by von Mises stress.}
	\label{figs:2D_rings_v0.12}
\end{figure}

Fig. \ref{figs:2D_rings_v0.12} shows the evolution of particle configuration for the SPH-OG, SPH-OAS and the present SPH-ENOG when the initial velocity magnitude $v_0=0.06c_0$. 
Clearly, the SPH-OG suffers from serious hourglass and tensile instability at the beginning of the computation ($t=0.002$), and the calculation process can barely continue. 
For the SPH-OAS, the tensile instability can be suppressed, and the particle distribution is uniform at the initial stage (t=0.002). 
However, with the passage of time, the zigzag distribution of particle configuration and von Mises stress gradually becomes apparent.
While for the SPH-ENOG, the particle and stress distribution are uniform during the whole calculation process, and the hourglass and tensile instability issues can be completely removed.

\begin{figure}[htb!]
	\centering
	\includegraphics[trim = 0cm 0cm 0cm 0cm, clip,width=0.95\textwidth]{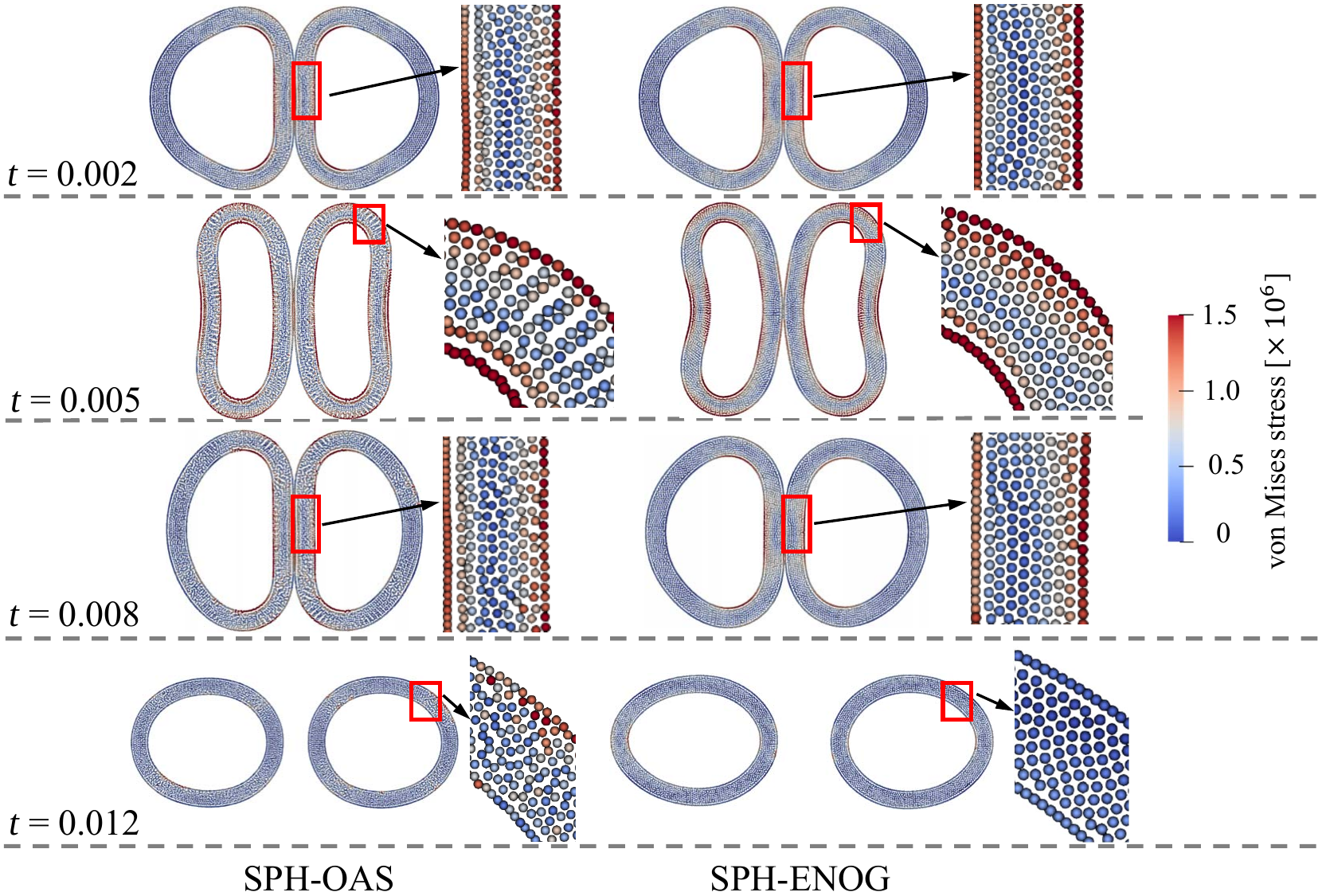}
	\caption{Evolution of particle configuration with time ($t=0.002$, 0.005, 0.008 and 0.012) for 2D colliding rubber rings. The results are obtained by different SPH methods, i.e., SPH-OAS (left column) and SPH-ENOG (right column). The initial velocity $v_0=0.07c_0$ and the figures are colored by von Mises stress.}
	\label{figs:2D_rings_v0.14}
\end{figure}

Then we increase the initial velocity $v_0$ to $0.07c_0$, to test the stability and robustness of the present SPH-ENOG. 
As shown in Fig. \ref{figs:2D_rings_v0.14}, the hourglass and tensile instability become more significant for the SPH-OAS, compared with Fig. \ref{figs:2D_rings_v0.12} when $v_0=0.06c_0$. 
Particularly, after the collision and rebound of the two rings ($t=0.012$), the zigzag distribution of  particles cannot be restored, indicating that this accumulated integration error over time is significant.
On the contrary, the present SPH-ENOG can produce a uniform particle and stress distribution throughout the entire process.

\begin{figure}[htb!]
	\centering
	\includegraphics[trim = 0cm 0cm 0cm 0cm, clip,width=0.95\textwidth]{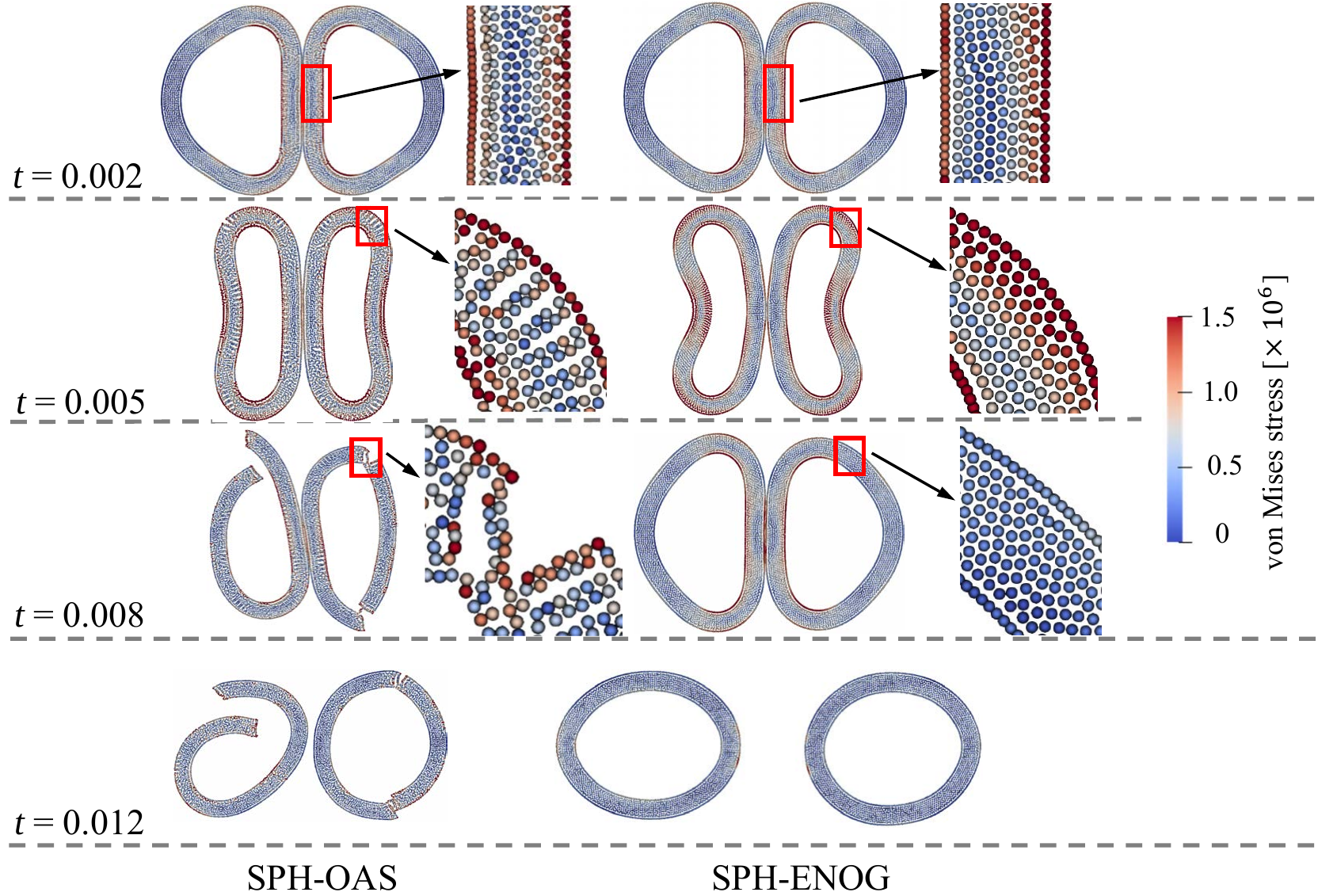}
	\caption{Evolution of particle configuration with time ($t=0.002$, 0.005, 0.008 and 0.012) for 2D colliding rubber rings. The results are obtained by different SPH methods, i.e., SPH-OAS (left column) and SPH-ENOG (right column). The initial velocity $v_0=0.08c_0$ and the figures are colored by von Mises stress.}
	\label{figs:2D_rings_v0.16}
\end{figure}

Furthermore, the initial velocity is increased to $v_0=0.08c_0$. It can be seen from Fig. \ref{figs:2D_rings_v0.16}, not only the hourglass, but also the tensile instability appears when $t\geqslant 0.005$ for the SPH-OAS. Fortunately, the present SPH-ENOG performs well even at such large initial velocity, and the hourglass and tensile instability can be perfectly eliminated, which suggests the stability and robustness of the present SPH-ENOG.

%%%%%%%%%%%%%%%%%%%%%%%%%%%%%%%%%%%%%%%%%%%%%%%%%%%%%%%%%%%%%
% 6.4 3D colliding rubber balls
%%%%%%%%%%%%%%%%%%%%%%%%%%%%%%%%%%%%%%%%%%%%%%%%%%%%%%%%%%%%%
\subsection{3D colliding rubber balls}\label{rubber-balls}

The 2D colliding rubber rings are extended to 3D to validate the proposed SPH-ENOG for 3D scenarios. The initial setup follows Fig. \ref{figs:2D-ring-setup}, i.e., two hollow rubber balls are moving towards each other with inner radius 0.03 and outer radius 0.04. The initial distance between the centers of the two balls is 0.09 and the initial velocity magnitude for each ball is $v_0$. The initial particle spacing $dp=0.001$ and the uniform particle distribution at the beginning is realized by Yu et al.'s method \cite{yu2023level}. The selection of material parameters follows section \ref{rubber-rings}.

\begin{figure}[htb!]
	\centering
	\includegraphics[trim = 0cm 0cm 0cm 0cm, clip,width=0.95\textwidth]{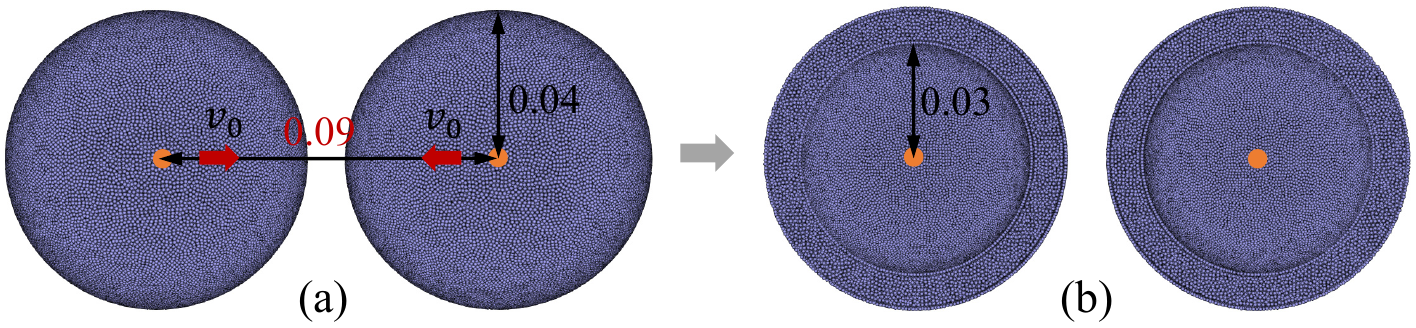}
	\caption{Model setup for (a) 3D colliding hollow rubber balls and (b) the half of each ball are showcased for proper visualization.}
	\label{figs:3D-balls-setup}
\end{figure}

\begin{figure}[htb!]
	\centering
	\includegraphics[trim = 0cm 0cm 0cm 0cm, clip,width=0.95\textwidth]{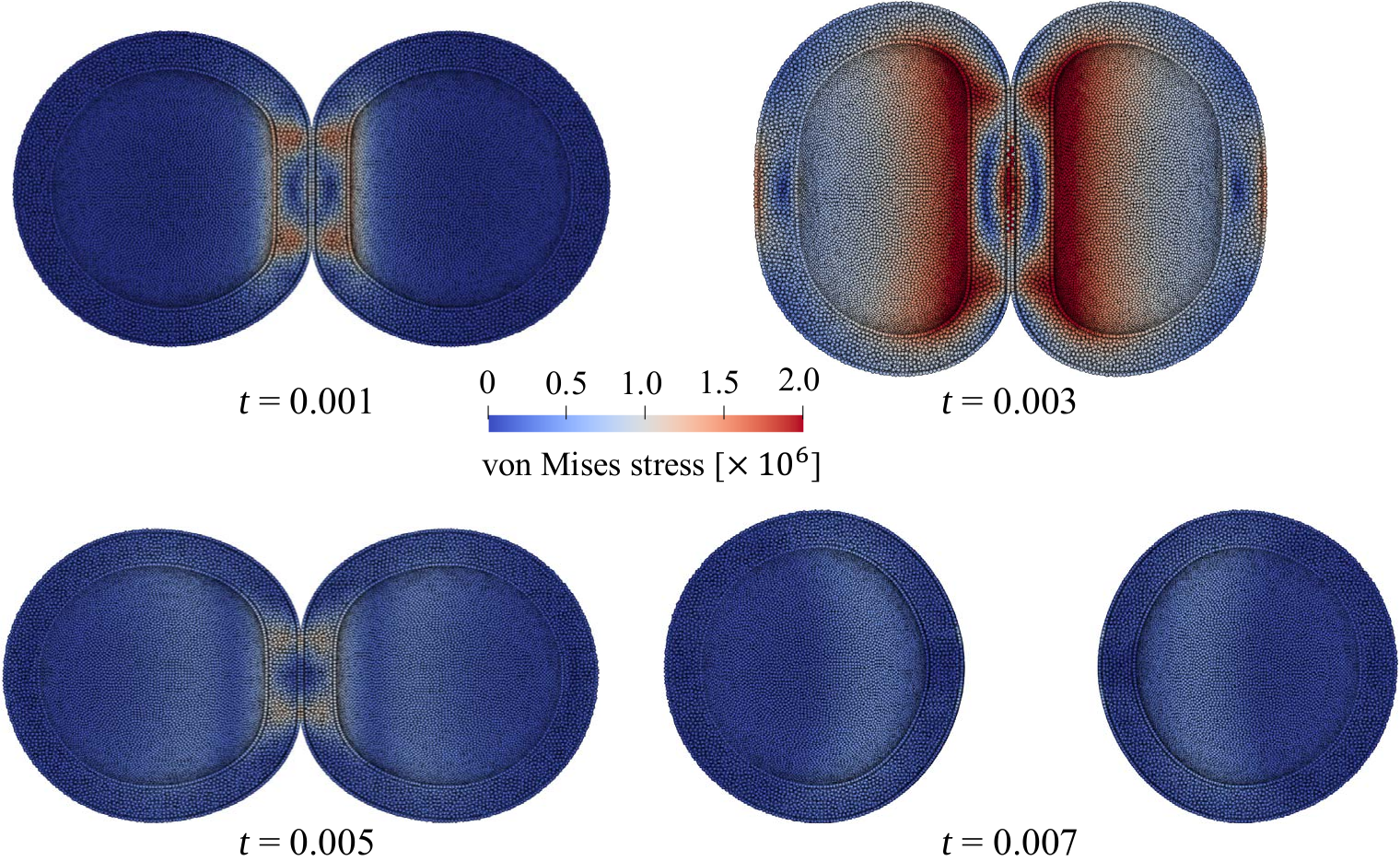}
	\caption{Evolution of particle configuration with time ($t=0.001$, 0.003, 0.005 and 0.007) for 3D colliding rubber balls. The results are obtained by the present SPH-ENOG. The initial velocity magnitude $v_0=0.08c_0$ and the figures are colored by von Mises stress.}
	\label{figs:3D-balls-0.16c0}
\end{figure}

Fig. \ref{figs:3D-balls-0.16c0} shows the collision process of two balls at different times ($t=0.001$, 0.003, 0.005 and 0.007) with the present SPH-ENOG. 
The initial velocity is set as $v_0=0.08c_0$. Half of each ball is displayed separately here for proper visualization. 
Obviously, the distribution of von Mises stress is smooth, and the particle configuration is uniform, which means hourglass modes and tensile instability can be completely removed. 

%%%%%%%%%%%%%%%%%%%%%%%%%%%%%%%%%%%%%%%%%%%%%%%%%%%%%%%%%%%%%
% 6.5 2D rubber ball-plate interaction
%%%%%%%%%%%%%%%%%%%%%%%%%%%%%%%%%%%%%%%%%%%%%%%%%%%%%%%%%%%%%
\subsection{2D rubber ball-plate interaction}
\label{2D-ball-plate}

\begin{figure}[htb!]
	\centering
	\includegraphics[trim = 0.5cm 0.6cm 0.5cm 0.2cm, clip,width=0.95\textwidth]{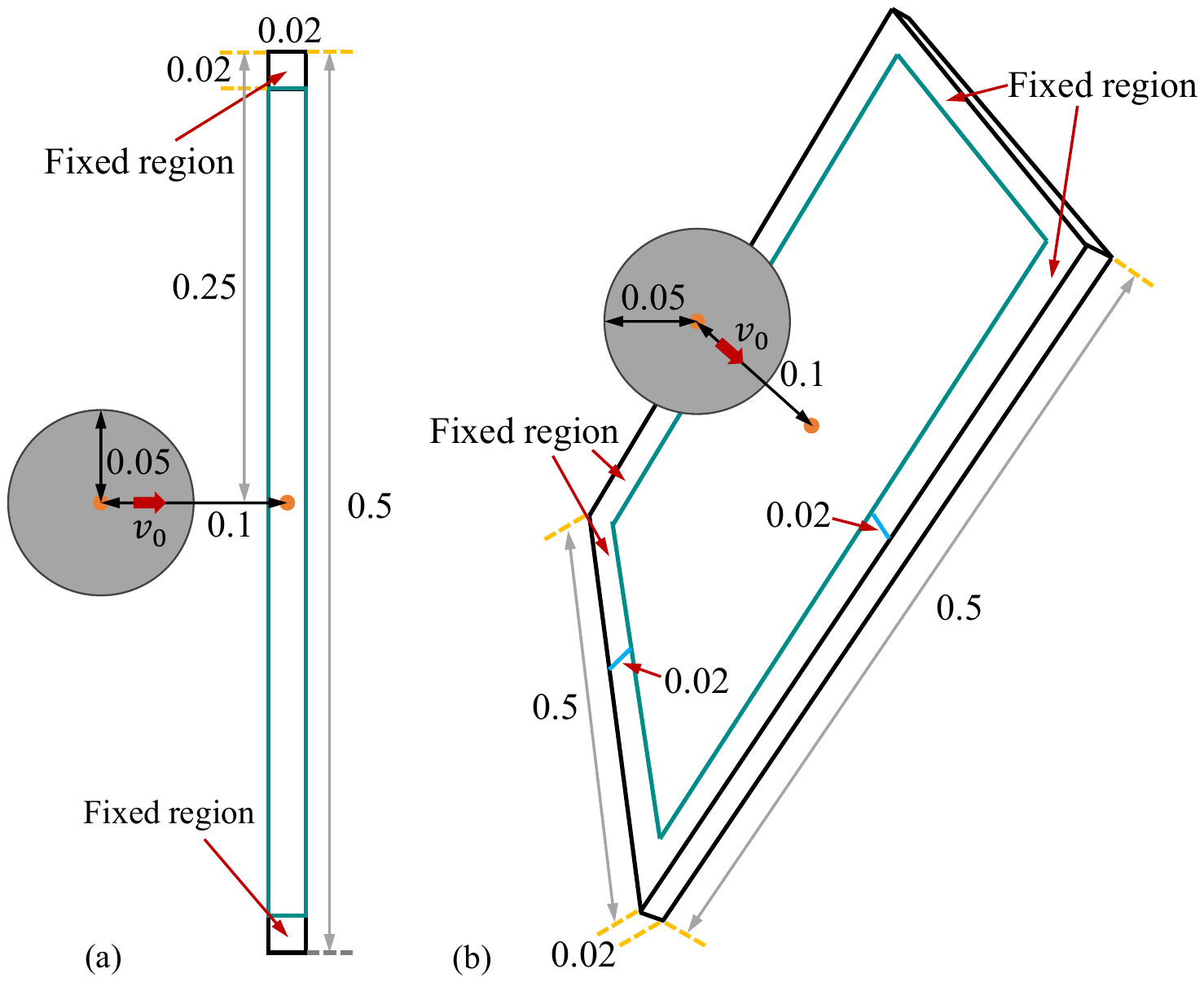}
	\caption{Model setup for (a) 2D rubber ball-plate interaction, and (b) 3D rubber ball-plate interaction.}
	\label{figs:2D-3D-ball-plate-setup}
\end{figure}

Another interesting problem with large deformation and tension produced is simulated in this section.
As shown in Fig. \ref{figs:2D-3D-ball-plate-setup}a, a rubber ball with a radius of 0.05 is used to impact a rubber plate \cite{zhang2017generalized}. The plate, fixed at both ends, has dimensions of 0.5 in length and 0.02 in width. The center of the ball is located 0.1 units away from the center of the plate, and the initial velocity of the rubber ball is $v_0$. 
The material parameters of the rubber ball and the target plate are the same \cite{zhang2017generalized}, i.e.,  density $\rho_0=1200$, Young's modulus $E=1\times 10^7$, and Poisson's ratio $\nu$ is set as 0.49 to produce a large deformation, mimicking realistic rubber materials. 
The initial particle spacing is 0.0025.

\begin{figure}[htb!]
	\centering
	\includegraphics[trim = 0cm 0cm 0cm 0cm, clip,width=0.9\textwidth]{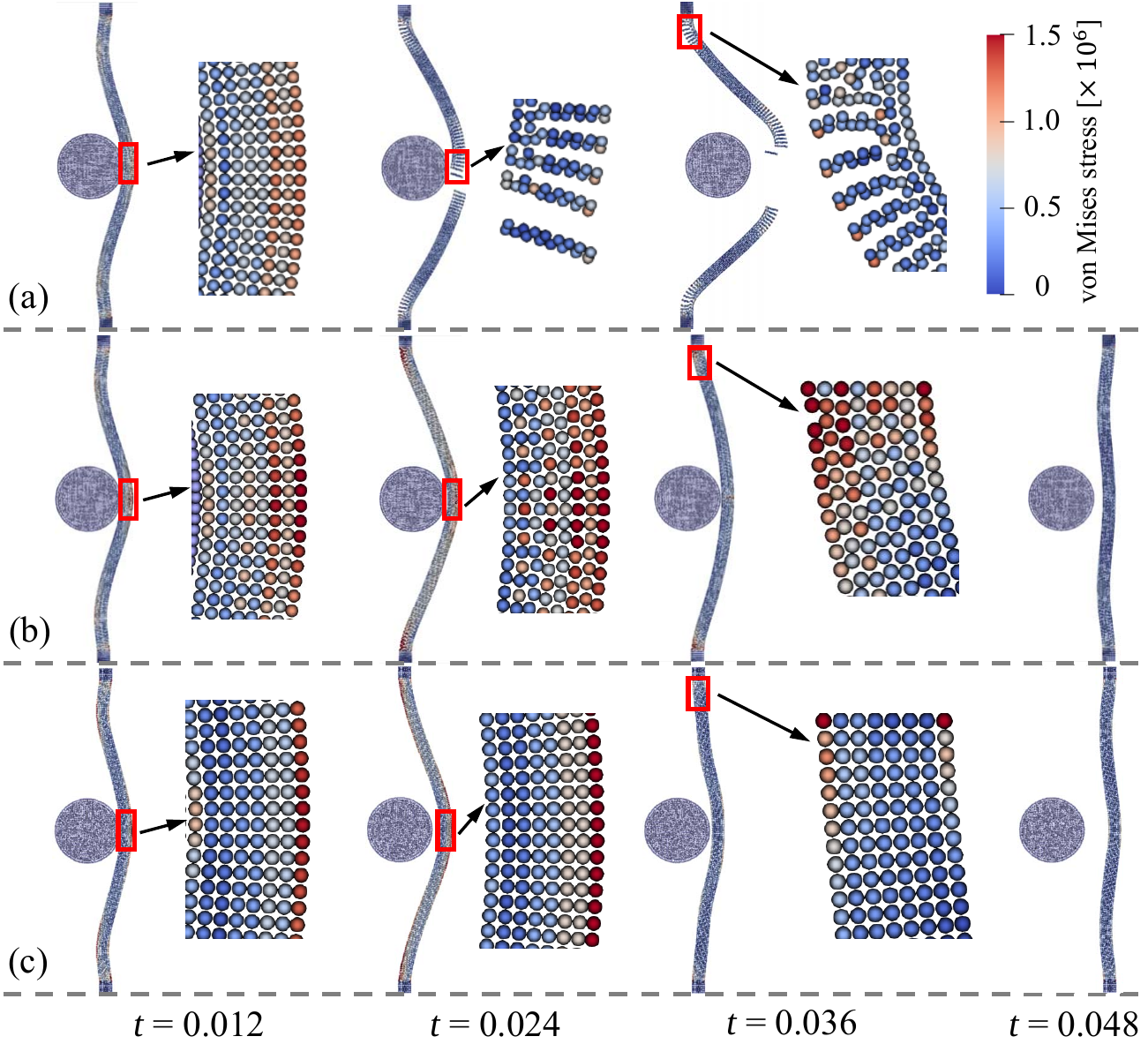}
	\caption{2D rubber ball-plate interaction with initial ball velocity 0.02$c_0$ at different times. The results are obtained by (a) SPH-OG, (b) SPH-OAS and (c) SPH-ENOG respectively. The figures are colored by von Mises stress.}
	\label{figs:2D-ball-plate-v0.02}
\end{figure}

Fig. \ref{figs:2D-ball-plate-v0.02} illustrates the process of collision between the rubber ball and the target plate with an initial velocity of 0.02$c_0$.  Significant tensions will be generated on the plate after the ball touches it. 
Serious hourglass modes (zigzag particle distribution) and tensile instability (numerical fractures) can be observed when using the SPH-OG (Fig. \ref{figs:2D-ball-plate-v0.02}a). For the SPH-OAS (Fig. \ref{figs:2D-ball-plate-v0.02}b), the tensile instability can be suppressed but the hourglass issue still exists. While with the present SPH-ENOG (Fig. \ref{figs:2D-ball-plate-v0.02}c), the tensile instability and hourglass modes can be removed simultaneously. The phenomena are consistent with the results from the oscillating plate and colliding rubber rings/balls described in Section \ref{2D-oscillating-plate} and Section \ref{rubber-rings}. 

\begin{figure}[htb!]
	\centering
	\includegraphics[trim = 0cm 0cm 0cm 0cm, clip,width=0.9\textwidth]{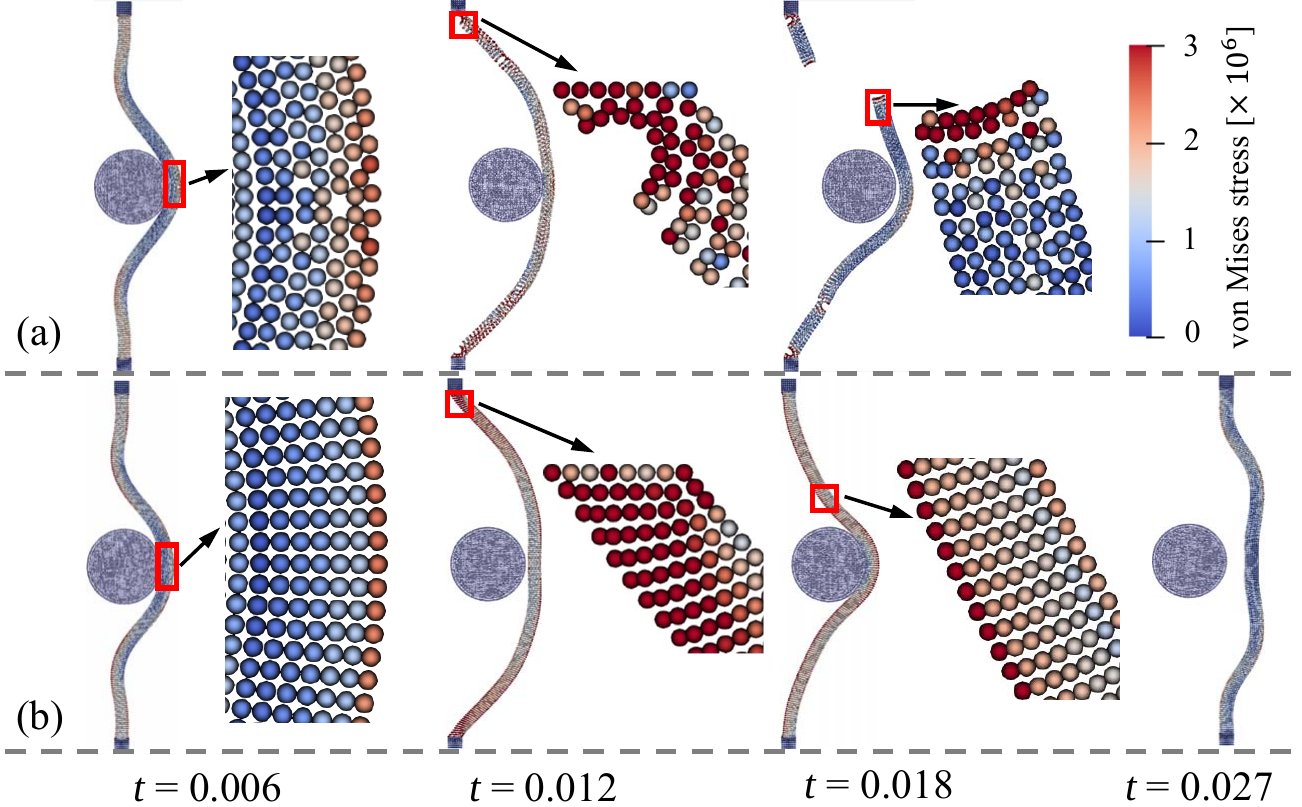}
	\caption{2D rubber ball-plate interaction with initial ball velocity 0.06$c_0$ at different times. The results are obtained by (a) SPH-OAS and (b) SPH-ENOG respectively. The figures are colored by von Mises stress.}
	\label{figs:2D-ball-plate-v0.06}
\end{figure}

Afterwards, we increase the initial velocity of the ball ($v_0=0.06c_0$) to test the stability and applicability of the SPH-OAS and SPH-ENOG under a  more demanding condition. 
It can be seen that the hourglass issue still exists for the SPH-OAS(Fig. \ref{figs:2D-ball-plate-v0.06}a), and to make matters worse, the tensile instability starts to appear at $t=0.012$ and become very severe later. 
However, our method (Fig. \ref{figs:2D-ball-plate-v0.06}b) still performs well under such condition, and both hourglass modes and tension instability do not occur.

\begin{figure}[htb!]
	\centering
	\includegraphics[trim = 0cm 0cm 0cm 0cm, clip,width=0.92\textwidth]{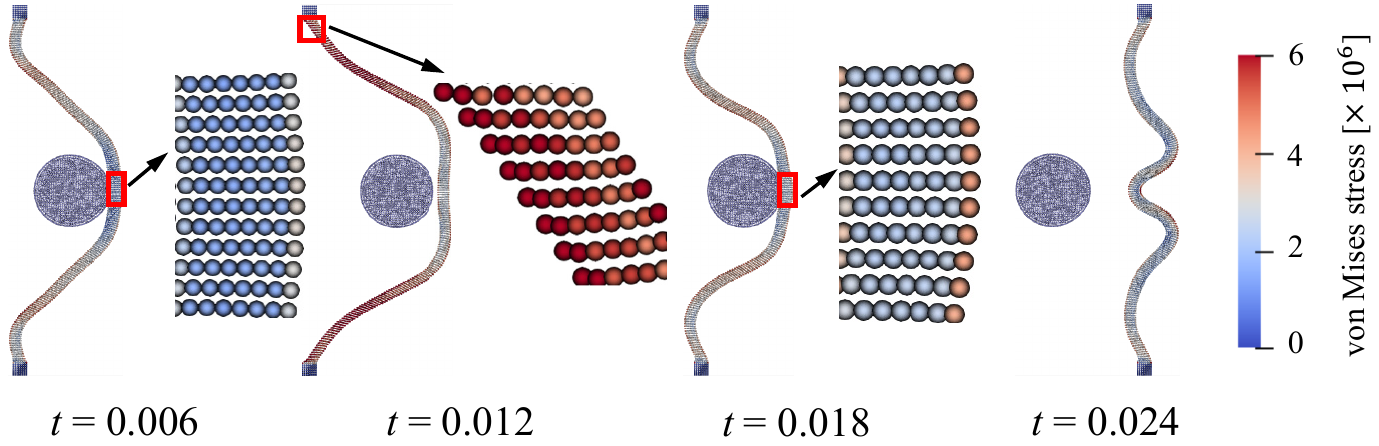}
	\caption{2D rubber ball-plate interaction with initial ball velocity 0.12$c_0$ at different times. The results are obtained by the present SPH-ENOG and the figures are colored by von Mises stress.}
	\label{figs:2D-ball-plate-v0.12}
\end{figure}

To further challenge the proposed SPH-ENOG, the initial velocity is set to $v_0=0.12c_0$ and extremely large tensions will be generated under such condition. As shown in Fig. \ref{figs:2D-ball-plate-v0.12},
surprisingly, the particle configuration still keeps a uniform distribution and a smooth stress profile can be obtained, implying the robustness and stability of the current SPH-ENOG.

%%%%%%%%%%%%%%%%%%%%%%%%%%%%%%%%%%%%%%%%%%%%%%%%%%%%%%%%%%%%%
% 6.6 3D sphere-rubber plate interaction
%%%%%%%%%%%%%%%%%%%%%%%%%%%%%%%%%%%%%%%%%%%%%%%%%%%%%%%%%%%%%
\subsection{3D rubber ball-plate interaction}
\label{3D-ball-plate}

We further consider the collision of the rubber ball and plate in 3D situations, as shown in Fig. \ref{figs:2D-3D-ball-plate-setup}b. The plate with a size of $0.5\times 0.02\times 0.5 $ is fixed on all four edges. The line connecting the center of the ball and the center of the plate is perpendicular to the plane of the plate, and the direction of initial velocity $v_0$ lies along this line. The materials of the rubber ball and plate follows section \ref{2D-ball-plate}, and the initial particle spacing is 0.0025.

\begin{figure}[htb!]
	\centering
	\includegraphics[trim = 0cm 0cm 0cm 0cm, clip,width=0.95\textwidth]{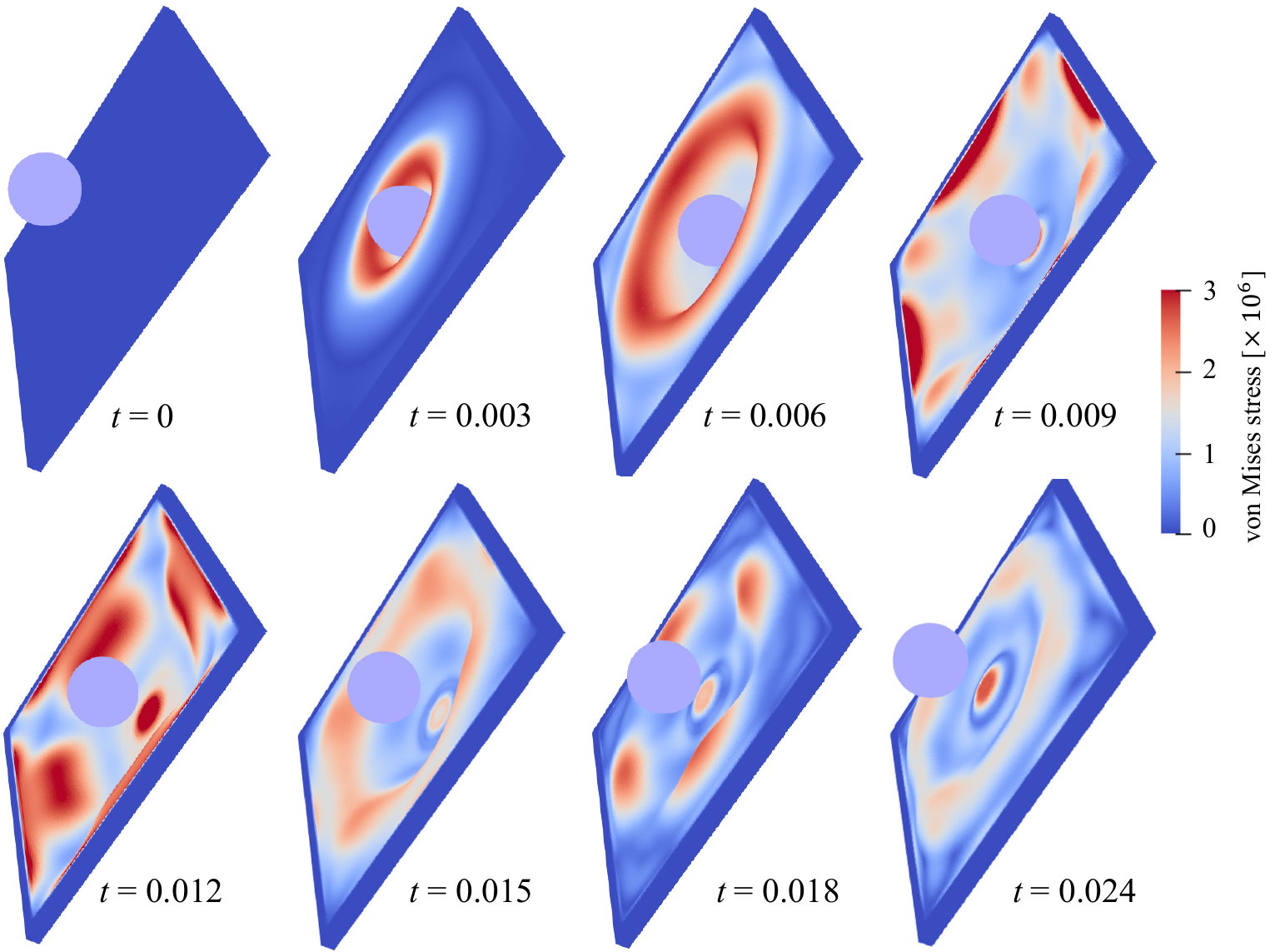}
	\caption{3D rubber ball-plate interaction with initial ball velocity 0.12$c_0$ at different times. The results are obtained by the present SPH-ENOG and the figures are colored by von Mises stress.}
	\label{figs:3D-ball-plate-v0.12}
\end{figure}

The results with the SPH-ENOG and $v_0=0.12c_0$ are shown in Fig. \ref{figs:3D-ball-plate-v0.12}. The profile of von Mises stress is smooth throughout the calculation process and there are no occurrences of numerical fractures. This demonstrates the capability of the present SPH-ENOG in eliminating hourglass modes and tensile instability for 3D situations.

%%%%%%%%%%%%%%%%%%%%%%%%%%%%%%%%%%%%%%%%%%%%%%%%%%%%%%%%%%%%%
%
% 7 Conclusion remarks
%
%%%%%%%%%%%%%%%%%%%%%%%%%%%%%%%%%%%%%%%%%%%%%%%%%%%%%%%%%%%%%
\section{Conclusion remarks}
\label{conclusions}

This paper presents an essentially non-hourglass and non-tensile-instability formulation for ULSPH by decomposing the shear acceleration into an angular-momentum conservative form with the Laplacian operator.
This newly-developed method is applicable for both 2D and 3D scenarios  without introducing case-dependent tuning parameters.
The convergence and accuracy of the present method are verified through several fundamental test cases.
Furthermore, the stability and robustness of the non-hourglass and non-tensile-instability formulation are confirmed through long-term simulations and simulations under extreme conditions. Unlike previous methods that mitigate tension instability through post-compensation and corrections, the current approach essentially eliminates tension instability from a novel perspective, i.e., resolving the hourglass mode in ULSPH.

This research also corrects a long-standing misconception regarding the causes of tension instability, and clarifies its true origin. In the past, tension instability was believed to be caused by tensions in elastic dynamics. However, this research demonstrates that if hourglass modes are addressed at its root, tension instability is naturally eliminated, even in scenarios involving extremely large tensions (as shown in Fig. \ref{figs:2D-ball-plate-v0.12}).
This implies that tension is not the true source of tension instability in elastic dynamics; rather, it is the further development and exacerbation of hourglass modes in the tensile region that causes it.

Last but not least, a dual-criteria time stepping scheme is employed to increase the calculation efficiency. 
For 2D and 3D oscillating plates, the computational time is reduced to approximately one-half and one-third, respectively, compared to the original calculations using single-criteria time stepping approach.

It should be noted that, although the present formulation is proposed and validated for elastic dynamics, 
it is possible to extend the current method to plastic dynamics for modelling material fracture and failure, as our framework is developed in the ULSPH.

%%%%%%%%%%%%%%%%%%%%%%%%%%%%%%%%%%%%%%%%%%%%%%%%%%%%%%%%%%%%%
%
% Section
%
%%%%%%%%%%%%%%%%%%%%%%%%%%%%%%%%%%%%%%%%%%%%%%%%%%%%%%%%%%%%%
\section*{CRediT authorship contribution statement}
\addcontentsline{toc}{section}{CRediT}

Shuaihao Zhang: Conceptualization, Methodology, Investigation, Visualization, Validation, Formal analysis, Writing - original draft, Writing - review \& editing. 
Sérgio D.N. Lourenço: Supervision, Investigation, Writing - review \& editing. 
Dong Wu: Investigation, Methodology, Formal analysis, Writing - review \& editing. 
Chi Zhang: Methodology, Writing - review \& editing.
Xiangyu Hu: Supervision, Investigation, Methodology, Writing - review \& editing.

%%%%%%%%%%%%%%%%%%%%%%%%%%%%%%%%%%%%%%%%%%%%%%%%%%%%%%%%%%%%%
%
% Section
%
%%%%%%%%%%%%%%%%%%%%%%%%%%%%%%%%%%%%%%%%%%%%%%%%%%%%%%%%%%%%%
\section*{Declaration of competing interest}
\addcontentsline{toc}{section}{declaration-interest}

The authors declare that they have no known competing financial interests or personal relationships that could
have appeared to influence the work reported in this paper.

%%%%%%%%%%%%%%%%%%%%%%%%%%%%%%%%%%%%%%%%%%%%%%%%%%%%%%%%%%%%%
%
% Section
%
%%%%%%%%%%%%%%%%%%%%%%%%%%%%%%%%%%%%%%%%%%%%%%%%%%%%%%%%%%%%%
\section*{Data availability}
\addcontentsline{toc}{section}{data-availability}

The code and data are available on GitHub.

%%%%%%%%%%%%%%%%%%%%%%%%%%%%%%%%%%%%%%%%%%%%%%%%%%%%%%%%%%%%%
%
% Section
%
%%%%%%%%%%%%%%%%%%%%%%%%%%%%%%%%%%%%%%%%%%%%%%%%%%%%%%%%%%%%%
\section*{Acknowledgements}
\addcontentsline{toc}{section}{acknowledgement}

Sérgio D.N. Lourenço would like to express his gratitude to the Research Grants Council Hong Kong for their sponsorship of this research under a Collaborative Research Fund (C6006-20GF).
Dong Wu, Chi Zhang and Xiangyu Hu would like to express their gratitude to the German Research Foundation (DFG) for their sponsorship of this research under grant number DFG HU1527/12-4.
The computations were performed using research computing facilities offered by Information Technology Services, the University of Hong Kong.

%%%%%%%%%%%%%%%%%%%%%%%%%%%%%%%%%%%%%%%%%%%%%%%%%%%%%%%%%%%%%
%
% Section
%
%%%%%%%%%%%%%%%%%%%%%%%%%%%%%%%%%%%%%%%%%%%%%%%%%%%%%%%%%%%%%
%\section*{Appendix}
%\label{appendix}
%%%%%%%%%%%%%%%%%%%%%%%%%%%%%%%%%%%%%%%%%%%%%%%%%%%%%%%%%%%%%
% Section
%%%%%%%%%%%%%%%%%%%%%%%%%%%%%%%%%%%%%%%%%%%%%%%%%%%%%%%%%%%%%
%\subsection*{Appendix A : The pairwise splitting approach}\label{appendix A}

%%%%%%%%%%%%%%%%%%%%%%%%%%%%%%%%%%%%%%%%%%%%%%%%%%%%%%%%%%%%%
%
% Section
%
%%%%%%%%%%%%%%%%%%%%%%%%%%%%%%%%%%%%%%%%%%%%%%%%%%%%%%%%%%%%%
\bibliographystyle{elsarticle-num}
\bibliography{non-hourglass-UL}
%%%%%%%%%%%%%%%%%%%%%%%%%%%%%%%%%%%%%%%%%%%%%%%%%%%%%%%%%%%%%
%
% Section
%
%%%%%%%%%%%%%%%%%%%%%%%%%%%%%%%%%%%%%%%%%%%%%%%%%%%%%%%%%%%%%
\end{document}